\documentclass[twocolumn]{aastex63}
\usepackage{amsmath}

\renewcommand{\vec}[1]{\mbox{\boldmath$#1$}}

\newcommand{\Ndesign}{\ensuremath{{N_\mathrm{design}}}}
\newcommand{\Ndimx}{\ensuremath{{N_{\mathrm{dim},\vec{x}}}}}
\newcommand{\Ndimf}{\ensuremath{{N_{\mathrm{dim},\vec{f}}}}}

\newcommand{\Obhh}{\ensuremath{\Omega_\mathrm{b}h^2}}
\newcommand{\Omhh}{\ensuremath{\Omega_\mathrm{m}h^2}}
\newcommand{\Onuhh}{\ensuremath{\Omega_\nu h^2}}
\newcommand{\sig}{\ensuremath{\sigma_8}}
\newcommand{\LCDM}{\ensuremath{\Lambda\mathrm{CDM}}}
\newcommand{\wCDM}{\ensuremath{w\mathrm{CDM}}}
\newcommand{\sumMnu}{\ensuremath{\sum m_\nu}}
\newcommand{\wo}{\ensuremath{w_0}}
\newcommand{\wa}{\ensuremath{w_a}}
\newcommand{\Mtwoc}{\ensuremath{M_{200\mathrm{c}}}}
\newcommand{\Mtwom}{\ensuremath{M_{200\mathrm{m}}}}

\newcommand{\MiraTitanUniverse}{\textit{Mira-Titan Universe}}
\newcommand{\Aemulus}{\textsc{Aemulus}}
\newcommand{\DarkEmulator}{\textsc{Dark Emulator}}

\shorttitle{Mira-Titan Universe III: Halo Mass Function Emulation}
\shortauthors{Bocquet et al.}
\submitjournal{\apj}

\begin{document}

\title{The Mira-Titan Universe. III. Emulation of the Halo Mass Function}

\author[0000-0002-4900-805X]{Sebastian Bocquet}
\email{sebastian.bocquet@physik.lmu.de}
\affiliation{\ANLHEP}
\affiliation{\Munich}
\affiliation{\Origins}
\author[0000-0003-1468-8232]{Katrin Heitmann}
\affiliation{\ANLHEP}
\author[0000-0002-7832-0771]{Salman Habib}
\affiliation{\ANLHEP}
\affiliation{\ANLCPS}
\author{Earl Lawrence}
\affiliation{\LANL}
\author{Thomas Uram}
\affiliation{\ANLALCF}
\author{Nicholas Frontiere}
\affiliation{\ANLHEP}
\affiliation{\ANLCPS}
\author{Adrian Pope}
\affiliation{\ANLCPS}
\author{Hal Finkel}
\affiliation{\ANLALCF}
\def\ANLALCF{ALCF Division, Argonne National Laboratory, Lemont, IL 60439, USA}
\def\ANLCPS{CPS Division, Argonne National Laboratory, Lemont, IL 60439, USA}
\def\ANLHEP{HEP Division, Argonne National Laboratory, Lemont, IL 60439, USA}
\def\LANL{CCS-6, CCS Division, Los Alamos National Laboratory, Los Alamos, NM 87545, USA}
\def\Munich{Faculty of Physics, Ludwig-Maximilians-Universit\"{a}t, Scheinerstr. 1, 81679 Munich, Germany}
\def\Origins{Excellence Cluster ORIGINS, Boltzmannstr. 2, 85748 Garching, Germany}

\begin{abstract}
We construct an emulator for the halo mass function over group and cluster mass scales for a range of cosmologies, including the effects of dynamical dark energy and massive neutrinos. The emulator is based on the recently completed \textit{Mira-Titan Universe} suite of cosmological $N$-body simulations. The main set of simulations spans 111 cosmological models with 2.1~Gpc boxes. We extract halo catalogs in the redshift range $z=[0.0, 2.0]$ and for masses $M_{200\mathrm{c}}\geq 10^{13}M_\odot/h$. The emulator covers an 8-dimensional hypercube spanned by \{$\Omega_\mathrm{m}h^2$, $\Omega_\mathrm{b}h^2$, $\Omega_\nu h^2$, $\sigma_8$, $h$, $n_s$, $w_0$, $w_a$\}; spatial flatness is assumed. We obtain smooth halo mass functions by fitting piecewise second-order polynomials to the halo catalogs and employ Gaussian process regression to construct the emulator while keeping track of the statistical noise in the input halo catalogs and uncertainties in the regression process. For redshifts $z\lesssim1$, the typical emulator precision is better than $2\%$ for $10^{13}-10^{14} M_\odot/h$ and $<10\%$ for $M\simeq 10^{15}M_\odot/h$. For comparison, fitting functions using the traditional universal form for the halo mass function can be biased at up to 30\% at $M\simeq 10^{14}M_\odot/h$ for $z=0$. Our emulator is publicly available at \url{https://github.com/SebastianBocquet/MiraTitanHMFemulator}.
\end{abstract}


\section{Introduction}

Matter in the Universe is known to cluster in the form of localized, clumpy distributions called halos. The abundance of massive halos as a function of total mass (dark matter and baryons) and redshift -- the mass function -- depends sensitively on cosmological parameters. In the context of this paper, the term ``massive" refers to halo masses characteristic of galaxy clusters, a rich and well-studied class of objects, with a central place in modern cosmology~\citep{2004cgpc.symp.....M}. Theoretical predictions of the mass function on observationally relevant mass scales -- an essentially nonlinear quantity -- can now be carried out with good accuracy using large-scale cosmological simulations. Measurements of cluster-scale halo masses can then be used to probe the evolution of cosmic structure growth and to constrain cosmological parameters (\citealt{Holder2001ApJ...560L.111H}; for a review, see \citealt{Allen2011ARA&A..49..409A}). In practice, such halos are detected via the (localized) presence of galaxies~(\citealt{2000AJ....120.2148G}, \citealt{2014ApJ...785..104R}), gravitational lensing signatures \citep{Miyazaki2018PASJ...70S..27M}, and/or the presence of hot ionized gas contained within the deep gravitational potential wells characteristic of rich galaxy groups and galaxy clusters (via X-ray emission (\citealt{2001A&A...369..826B}, \citealt{2007ApJS..172..561B}, \citealt{2016A&A...592A...2P}) and Sunyaev-Zel'dovich effect observations (\citealt{2015ApJS..216...27B}, \citealt{PlanckCollaboration2016A&A...594A..27P}, \citealt{Hilton2018ApJS..235...20H})).
The cosmological constraints resulting from the analyses of cluster samples already allow for competitive measurements of the growth history of the universe \citep{Mantz2015MNRAS.446.2205M, Bocquet2019ApJ...878...55B, Costanzi2019MNRAS.488.4779C, Zubeldia2019MNRAS.489..401Z}. In the near future, large samples of thousands of clusters, combined with a mass calibration that is accurate at the few-percent level will allow for greatly improved measurements of the cluster mass function \citep[e.g.,][]{DES2020clusterabundance}.

From the theoretical perspective, halos suffer from the lack of a strict definition, and therefore, the notion of ``halo mass'' inherits a certain ambiguity~(see, e.g., \citealt{2001A&A...367...27W}). Aside from this fact, the halo mass itself is not a direct observable. For these reasons alone, connecting observations to a theoretically obtained mass function is a nontrivial task. In the absence of a robust forward modeling approach based on detailed simulations, the current state-of-the-art relies on empirical ``mass--observable'' relations to connect theory and observations. In such an approach, one important limitation is the accuracy with which the theoretical mass function is known (for a given definition of mass), as the relevant cosmological parameters are varied. This general topic is the focus of the work presented here.

Early estimates of the mass function applied the spherical collapse model to the linear matter density field~\citep{Press1974ApJ...187..425P}, further formalized via the excursion set approach \citep{Bond1991ApJ...379..440B}. In this approximate methodology, all dependence of the mass function on redshift and cosmology is completely described by the RMS fluctuations $\sigma(M,z)$ in the linear matter power spectrum $P(k,z)$. It is not obvious, however, that this ``universal'' prediction of the mass function would be sufficiently accurate. Initial results from numerical studies of the mass function based on $N$-body simulations found that this universality did hold at an approximate level, given a certain halo mass definition \citep{Jenkins2001MNRAS.321..372J}. A number of fits for the mass function have since been derived from $N$-body simulations, based on the idea that the universal form can be applied to multiple cosmological models across a wide range of redshifts~\citep[see, e.g.,][]{Sheth1999MNRAS.308..119S, Jenkins2001MNRAS.321..372J, Springel2005Natur.435..629S, Warren2006ApJ...646..881W, Heitmann2006ApJ...642L..85H}. (For a discussion on the possible halo definition-dependence of the universal form, see \citealt{White2002ApJS..143..241W}.)

Despite the remarkable success of the approach described above, as simulation results were further refined it was discovered that the redshift evolution of the mass function, even for $\Lambda$CDM, deviates from the universal prediction (at the $5-10\%$ level) and that this deviation had to be explicitly fitted \citep[e.g.,][]{Reed2007MNRAS.374....2R, Lukic2007ApJ...671.1160L, Cohn2008MNRAS.385.2025C, Tinker2008ApJ...688..709T, Crocce2010MNRAS.403.1353C, Bhattacharya2011ApJ...732..122B, Courtin2011MNRAS.410.1911C}.
Furthermore, it was found that the universal mass function fit established for a reference $\Lambda$CDM model can only be extrapolated to $w$CDM models (with $w\simeq-1$) at about 10\% accuracy \citep[e.g.,][]{Bhattacharya2011ApJ...732..122B}.

In order to obtain precision cosmological constraints from cluster samples, one therefore needs to proceed beyond the use of fitting functions of the type discussed above, given their limitations in accuracy and parametric coverage. Reducing the systematic uncertainty in the theoretical prediction by roughly an order of magnitude is not an easy task \citep[for a detailed discussion of these issues, see, e.g.,][]{Bhattacharya2011ApJ...732..122B}, especially if one wishes to include the effects due to non-zero neutrino mass and dynamical dark energy, and also account for the influence of baryons, all of which have potential ramifications for the behavior of the mass function.

In the cosmological context, a direct numerical approach to this problem with a finite number of sufficiently accurate simulations is in fact possible using a combination of efficient sampling strategies, Bayesian statistical methods, and machine-learning based data reduction and interpolation; a process termed {\emph{emulation}}~\citep[e.g.,][]{Heitmann2006ApJ...646L...1H, Habib2007PhRvD..76h3503H, Higdon2010}. The end result of the emulation process, an {\em emulator}, is an oracle that, given a set of input parameters, yields an essentially instantaneous prediction for a set of summary statistics, with well-defined errors. 
 
To construct an emulator, numerical simulations are first run for a set of cosmologies sampling a bounded parameter space. Statistical data reduction and machine-learning based interpolation techniques then yield the desired predictions for observables for any set of parameters contained within the sampled region. Emulators have been used to successfully predict the mass function and other nonlinear summary statistics characteristic of cosmological structure formation such as the nonlinear matter power spectrum, halo bias, and halo concentration \citep{Lawrence2010ApJ...713.1322L, Kwan2013ApJ...768..123K, Heitmann2014ApJ...780..111H, Kwan2015ApJ...810...35K, Lawrence2017ApJ...847...50L, Nishimichi2019ApJ...884...29N, McClintock2019ApJ...872...53M}.

In this paper, we present a mass function emulator using the \MiraTitanUniverse\ suite of $N$-body simulations~\citep{Heitmann2016ApJ...820..108H}. This simulation suite includes the effects of massive neutrinos as well as dynamical dark energy and is therefore suited for current and next-generation cosmological surveys that are searching for deviations from $\Lambda$CDM. The box sizes (2.1~Gpc) and mass resolution ($3200^3$ particles) of the simulations were designed to measure the mass function at mass scales starting at $\sim10^{13}~M_\odot/h$ with good resolution of individual halos (at least 1000 particles per halo) as well as good statistics of halo masses in individual mass bins on group/cluster mass scales. Our emulator provides the mass function up to redshift $z=2$, with percent-level accuracy at group-scale masses and better than $\sim10\%$ accuracy at $10^{15}M_\odot/h$. This new emulator is a complement to the Mira-Titan emulator for the matter power spectrum presented in \cite{Lawrence2017ApJ...847...50L}.

This paper is structured as follows. The \MiraTitanUniverse\ simulations and the extraction of binned halo catalogs are discussed in Section~\ref{sec:simulations}. We describe the construction of the halo mass function emulator and verify its accuracy in Section~\ref{sec:emu_construction}. We discuss existing predictions for the mass function in Section~\ref{sec:literatureHMF}. In Section~\ref{sec:universality}, we use the \MiraTitanUniverse\ simulations to discuss the limits of cosmological universality of the mass function. We conclude with a summary and outlook in Section~\ref{sec:summary}.


\section{The Mira-Titan Universe}
\label{sec:simulations}

 The simulations used in this work were run using the HACC cosmological $N$-body code~\citep[Hardware/Hybrid Accelerated Cosmology Code,][]{Habib2016NewA...42...49H}. The suite of simulations is named the \MiraTitanUniverse\ after the supercomputers used to produce it (the IBM BlueGene/Q system Mira at Argonne and the GPU-accelerated system Titan at Oak Ridge) and is introduced in~\cite{Heitmann2016ApJ...820..108H}. Here, we summarize the relevant aspects, and refer the reader to the original paper and references therein for further details.

\subsection{Cosmological Modeling: $\nu w_0 w_a$CDM}

The \MiraTitanUniverse\ is a suite of cosmological $N$-body simulations that are realizations of 111 different cosmologies, which we refer to as M001--M111. The cosmological parameters are drawn from an eight-dimensional parameter space \{\Omhh, \Obhh, \Onuhh, \sig, $h$, $n_s$, \wo, \wa\}. All models are spatially flat ($\Omega_k=0$).
We consider a dynamical dark energy equation of state, using the common parameterization \citep{2001IJMPD..10..213C, Linder2003PhRvL..90i1301L}
\begin{equation} \label{eq:w0wa}
w(a) = w_0 + w_a(1-a).
\end{equation}
We further define the parameter combination
\begin{equation}
w_b \equiv (-w_0-w_a)^{1/4}
\end{equation}
as we will not consider the ($w_0,w_a$) parameter space but ($w_0,w_b$) (see discussion after Eq.~\ref{eq:wb}).

Massive neutrinos are treated at the background level instead of being simulated as a separate particle species, essentially an expansion in the neutrino mass fraction, $f_{\nu}$, in the spirit of \cite{Saito2009PhRvD..80h3528S}. A detailed description of our approach is given in \cite{Upadhye2014PhRvD..89j3515U} and \cite{Heitmann2016ApJ...820..108H}. Its validity with regard to power spectrum measurements on large to quasi-linear scales is discussed extensively in~\cite{Upadhye2014PhRvD..89j3515U} and the effect of neutrinos on the mass function is investigated for two models in \cite{Biswas2019arXiv190110690B}, including one for high-mass neutrinos. For completeness, we provide a short summary of the approach here. As mentioned above, we do not include the nonlinear evolution of the neutrinos explicitly in the simulation but rather evolve the cold dark-matter-baryon component only. The neutrinos are included in the background evolution (and the initial condition) and therefore do affect matter clustering in the nonlinear regime. However, the neutrino clustering itself is not taken into account. Particular care is given to the set-up of the initial conditions. We include the neutrino contribution in the transfer function and generate a linear power spectrum at $z=0$ with the chosen $\sigma_8$. We then move the linear power spectrum back to the initial redshift using the scale-independent growth function (note that the growth function including the nonlinear neutrino evolution is scale-dependent.) The growth function includes all species in the homogeneous background. The use of the scale-independent growth function is important since the evolution is carried out only for the dark-matter-baryon component. This approach ensures that the fully developed total (clustered) matter power spectrum at redshift $z=0$ is correctly normalized to a given $\sigma_8$. In this approximate approach the linear fluctuations in the neutrinos have to be added separately to get the power spectrum, where the nonlinear contribution of the neutrinos is neglected. At the current observationally favored neutrino mass range, $\sim 0.1$~eV or less, the nonlinear clustering of neutrinos has a negligible effect on halo masses, while at the very upper end of the masses considered here ($\sim 1$~eV), the effect on cluster-scale halos is sub-dominant compared to the overall suppression of the mass function due to neutrino free-streaming, and is of the order of the overall accuracy of the emulator.

The simulation suite used in this work covers gravity-only simulations. The expectation is that direct modeling of baryonic effects (as well as approaches based on post-processing gravity-only simulations) will be added over time~\citep{Heitmann2016ApJ...820..108H}.

\begin{deluxetable*}{lccccc}
\tablecaption{\label{tab:sims}
\MiraTitanUniverse\ simulations used to construct the emulator.}
\tablehead{\colhead{Model} & \colhead{Box size} & \colhead{$N_\mathrm{particle}$} & \colhead{Force res.}
& \colhead{$M_\mathrm{particle}$} & \colhead{$\mathrm{min}(\Mtwoc)$}}
\startdata
M001--M111 & $2.100\,\mathrm{Gpc}$ & $3200^3$ & 6.6 kpc & $1.08\times10^{10}\left(\frac{\Omega_\mathrm m - \Omega_\nu}{0.28}\right)\left(\frac{h}{0.7}\right)^2M_\odot$ & $10^{13}M_\odot/h$ \\\tableline
M006 & $2.091\,\mathrm{Gpc}$ & $3200^3$ & 6.6 kpc & $1.02\times10^{10}M_\odot$ & $10^{13}M_\odot/h$ \\
M023 & $2.085\,\mathrm{Gpc}$ & $3200^3$ & 6.6 kpc & $9.96\times10^{9}M_\odot$ & $10^{13}M_\odot/h$ \\
M046 & $1.865\,\mathrm{Gpc}$ & $3200^3$ & 6.6 kpc & $7.23\times10^{9}M_\odot$ & $10^{13}M_\odot/h$
\enddata
\tablecomments{The mass cuts $\mathrm{min}(\Mtwoc)$ ensures we only use well-resolved halos with $N_\mathrm{particle}>1000$. The simulations for models M006, M023, and M046 were accidentally run with slightly smaller box sizes than 2.1~Gpc. As we build an emulator for the spatial halo number \emph{density} this change in volume is trivial to account for.}
\end{deluxetable*}

\subsection{The Design}

The \MiraTitanUniverse\ was specifically designed to produce simulation data for emulators. The volume of the sampled parameter space is therefore a compromise between covering wide parameter ranges and the sparsity this implies. The sampling design underlying the \MiraTitanUniverse\ has an inbuilt notion of sequential convergence and is aimed at generating multiple emulators at percent-level error levels, including the halo mass function \citep[as demonstrated in a first test in][]{Heitmann2016ApJ...820..108H}.

\begin{figure*}[ht]
  \includegraphics[width=\textwidth]{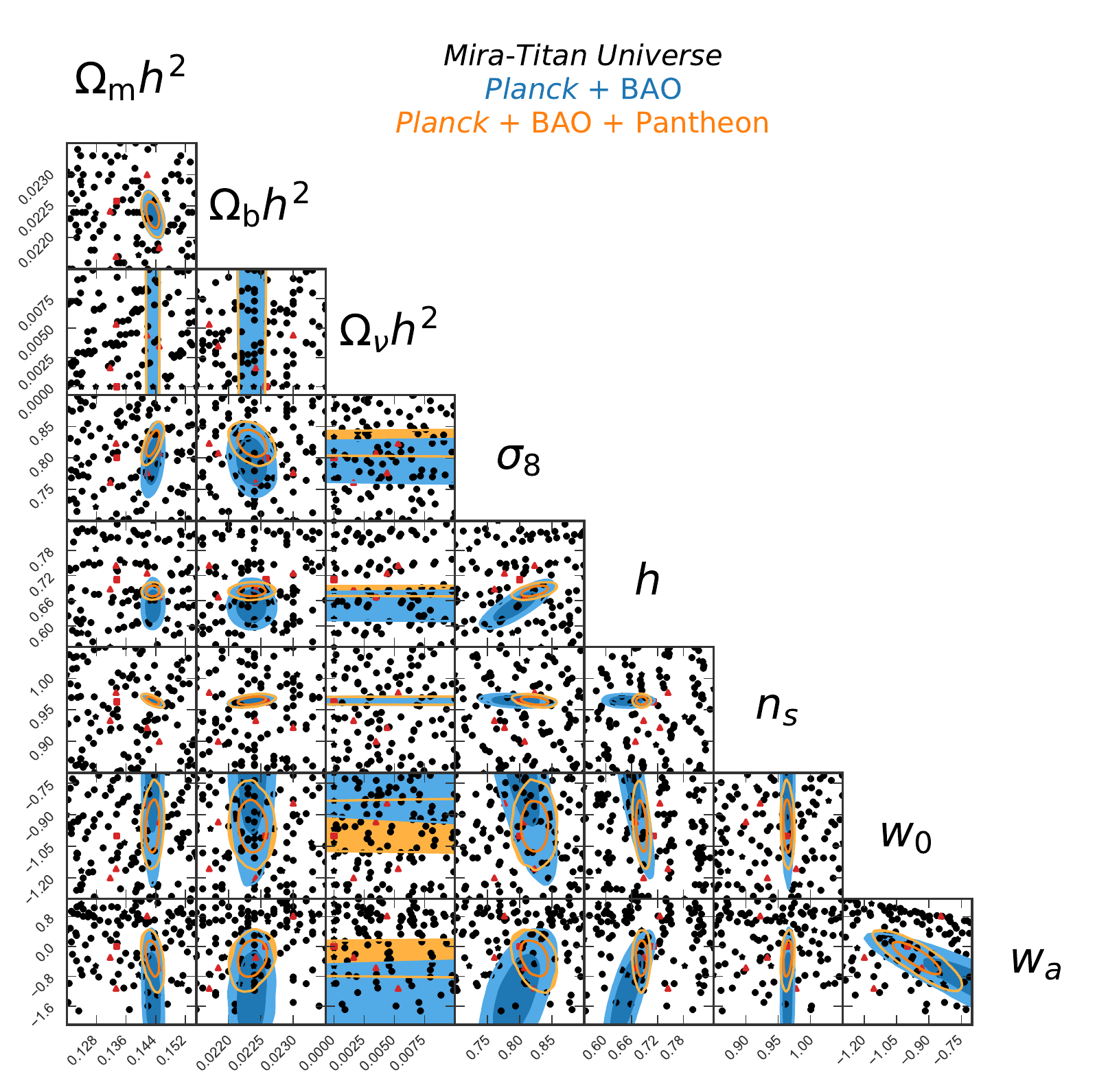}
  \caption{Cosmologies of the \MiraTitanUniverse\ (111 black markers). Black stars show the models M001--M010 with massless neutrinos, black dots show the models M011--M111 with massive neutrinos. The red boxes show the M000 cosmology (massless neutrinos, cosmological constant) and the red triangles show the remaining four test models T001 through T004. 
  Colored contours show the constraints from \textit{Planck}+BAO and \textit{Planck}+BAO+Pantheon.}
  \label{fig:GTC}
\end{figure*}

The cosmological parameters are chosen within the following boundaries:
\begin{eqnarray}\label{eq:cosmoparams}
0.12\le &\Omega_\mathrm m h^2& \le 0.155\\
0.0215\le &\Omega_\mathrm b h^2& \le 0.0235\\
0.0\le &\Omega_\nu h^2& \le 0.01\\
0.7\le &\sigma_8& \le 0.9\\
0.55\le &h& \le 0.85\\
0.85\le &n_s& \le 1.05\\
-1.3\le &w_0& \le -0.7\\
0.3 \le &w_b& \le 1.3 \label{eq:wb}
\end{eqnarray}
The cosmological hypercube is spanned by $w_b$ instead of $w_a$ to ensure a better coverage of models with $w_0+w_a$ close to zero \citep{Heitmann2016ApJ...820..108H}. As a result, $w_a$ is jointly constrained with $w_0$; the smallest allowed value is $w_a=-2.16$ and the highest allowed value is $w_a=1.29$.
The range in \Onuhh\ corresponds to a range in the sum of neutrino masses $0\le\sumMnu\le0.94$~eV. While the upper limit is significantly higher than current results for a \LCDM\ background cosmology, the constraint on \Onuhh\ significantly degrades when $w_0$ and $w_a$ are also allowed to vary (see Figure~\ref{fig:GTC}).

The choice of the 111 \emph{design models} and their space-filling properties are discussed in Section~3 in~\cite{Heitmann2016ApJ...820..108H}. Note that a tessellation-based nested design strategy was specially developed in order to obtain sequential convergence, allowing reuse of previously run simulations: The first 26 models (M011--M036) represent a complete design in their own right, which is further refined when adding the next 29 models to obtain a design with 55 models. Adding yet another 46 models gives a full design with 101 models. To improve coverage of the edge of parameter space where $\sumMnu=0$, another set of 10 models (M001--M010) with massless neutrinos are chosen according to a seven-parameter symmetric latin hypercube design. All cosmologies in the Mira-Titan design are listed in Table~\ref{tab:cosmo} and are visualized in Figure~\ref{fig:GTC}.\footnote{Figure~\ref{fig:GTC} also shows current cosmological constraints from \textit{Planck} TTTEEE+lowl+lowE \citep{PlanckCollaboration2018arXiv180706209P}, Baryon Acoustic Oscillations \citep[BAO,][]{Beutler2011MNRAS.416.3017B, Ross2015MNRAS.449..835R, Alam2017MNRAS.470.2617A}, and the Pantheon supernova sample \citep{Scolnic2018ApJ...859..101S}.} As discussed earlier, the design hypercube is constructed using $w_0$ and $w_b$, which leads to the impression of uneven sampling in ($w_0,w_a$) space.

For each of the 111 design cosmologies, a 2.1~Gpc box simulation was run evolving $3200^3$ particles. The particle masses vary between $7.23\times10^9\,M_\odot$ and $1.22\times10^{10}\,M_\odot$ depending on the simulated cosmology, comfortably resolving group and cluster-scale halos. 
Key features of the simulations are summarized in Table~\ref{tab:sims}.

An important verification test of the emulator accuracy is to compare its mass function prediction directly with numerical results from additional (off-design) simulations that were not used for the construction of the emulator. For this purpose, a set of four additional 2.1~Gpc box simulations were run (T001--T004). Finally, another 2.1~Gpc simulation box was run using a fiducial $\Lambda$CDM cosmology (M000) that has also been used for other, larger simulations \citep[e.g., the Outer Rim simulation,][]{Heitmann2019ApJS..245...16H}. All of the five reference simulations evolve $3200^3$ particles as in our main simulation suite.


\begin{figure}
    \includegraphics[width=\columnwidth]{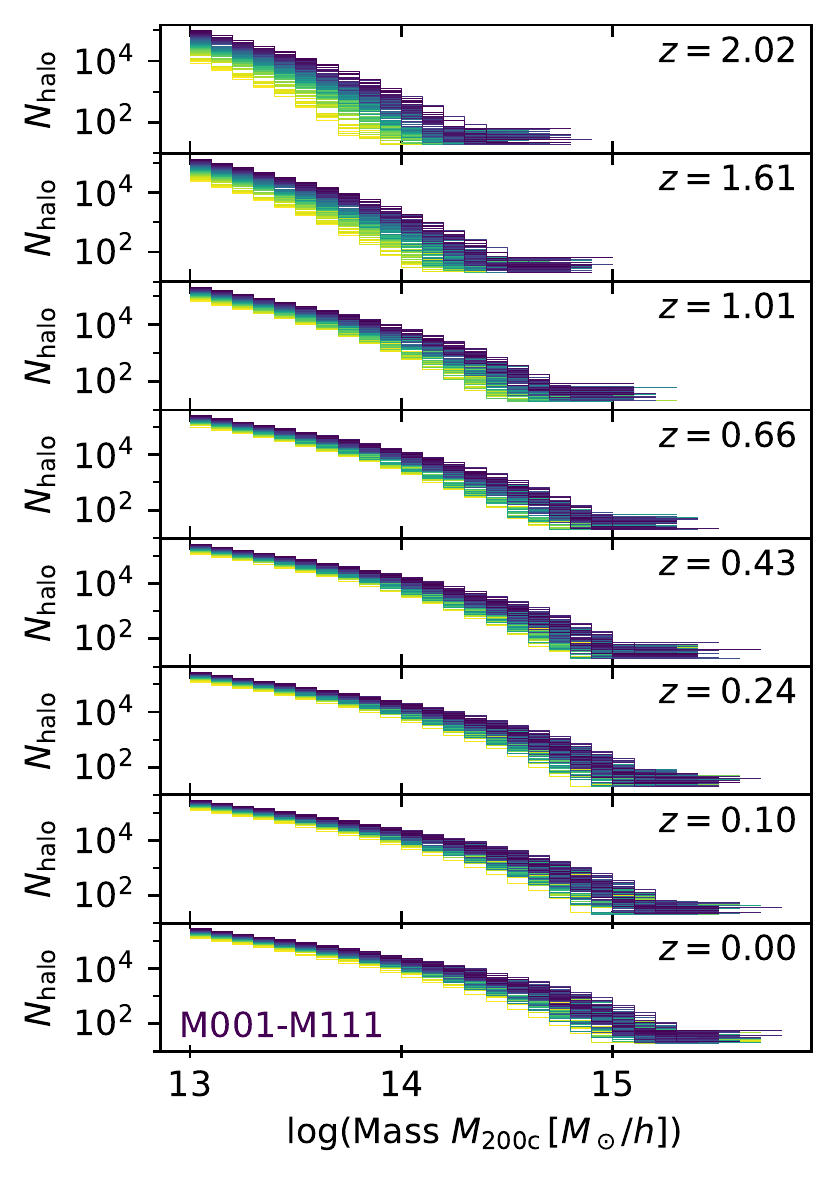}
    \caption{Binned halo catalogs from the 111 \MiraTitanUniverse\ simulations. The color gradient reflects the number of halos in the lowest mass bin. At high mass, we choose an adaptive binning scheme to ensure that each mass bin is populated with at least 20 halos.}
    \label{fig:design_HMF_1}
\end{figure}

\subsection{Halo Catalogs and Binned Halo Mass Function}
\label{sec:halo_catalogs}

We first identify halos using a fast parallel friends-of-friends (FOF) halo finder with link length set to $b=0.168$. Spherical overdensity (SO) halo catalogs are then built out from the potential minimum of each FOF halo. (In this second process, we use all of the local simulation particles, not just those found by the FOF algorithm.) SO halo masses are specified via the overdensity criterion $\Delta = 200 \rho_\mathrm{crit}$. We create halo catalogs for 8 snapshots that are roughly equally spaced in time within $0\le z\le2.02$. The redshifts are $z\in\text{\{0.00, 0.10, 0.24, 0.43, 0.66, 1.01, 1.61, 2.02\}}$. Throughout the simulation suite, we only consider halos above $10^{13}M_\odot/h$; this cut ensures that all halos are well resolved with at least 1000 particles.

For each halo catalog, we apply a binning in mass which is a compromise between good mass resolution at low mass where halos are most abundant and a sufficiently large number of halos per bin at high mass. First, we bin all halo catalogs with a default bin width $\Delta \log_{10}M = 0.1$. Then, we combine the highest-mass bins such that the resulting (wide) high-mass bin contains at least 20 halos.
Using $4^3$ spatial jackknife sub-volumes, we compute covariance matrices between mass bins to account for noise in the binned halo catalogs due to sample variance and shot noise. We checked that our mass function emulator is robust to changes in the details of the binning scheme and the estimation of the statistical noise. We ignore correlations between mass bins across different time snapshots. In Figure~\ref{fig:design_HMF_1}, we show the binned halo mass functions for the 111 design cosmologies.


\section{Emulator Construction}
\label{sec:emu_construction}

In this section, we discuss our emulator framework. We treat each snapshot separately, and thus construct 8 independent emulators. The key steps in building the emulators are summarized below; we discuss each of these steps in detail in the following subsections.
\begin{enumerate}
    \item For every input cosmology, obtain a smooth halo mass function model from the halo catalog 
    \item Perform a principal component analysis (PCA) on the (logarithm of the) smooth mass function models
    \item Keep the first four eigenvectors as basis functions
    \item Fit the four basis functions to each halo catalog to obtain a four-dimensional posterior parameter distribution for each model
    \item Set up Gaussian Process regression to interpolate between the four fit parameters, accounting for their covariances
    \item Set the hyperparameters of the Gaussian Process
    \item Verify the emulator predictions using hold-out tests and spot checks against additional simulations
\end{enumerate}
The emulator is now complete and, for a given requested cosmology, returns the halo mass function and an error estimate.

For our emulator and its construction, we consider mass in units of $M_\odot/h$ and volume in units of $(\mathrm{Mpc}/h)^3$. With this convention, the conversion from unit volume to the volume contained in a survey solid angle does not explicitly depend on the Hubble constant $h$.


\subsection{Constrained Piecewise Polynomial Halo Mass Function Fits}
\label{sec:HMFfit}

To ensure that the final emulator predictions for the mass function are smooth, we need to convert the binned halo catalogs into smooth functions of mass. Due to the large variance of the mass function at high mass, we found smoothing kernels not to be an adequate approach. Inspired by the approximate power-law behavior of the mass function at low masses, we instead fit a model consisting of constrained piecewise second-order polynomials in log-mass. In each segment $i$, we have
\begin{equation}
\label{eq:piecewiseHMF}
    \ln\left(\frac{dn}{d\ln M}\right)_i = a_i + b_i \ln\left(\frac{M}{M_\mathrm{piv}}\right) + c_i \ln\left(\frac{M}{M_\mathrm{piv}}\right)^2.
\end{equation}
We constrain the individual segments to connect smoothly by requiring the function and its derivative to be continuous. For every two consecutive segments joining at mass $M_\mathrm{join}$, this leads to a constraint on the function value
\begin{align}
\label{eq:constraintamp}
a_i &+ b_i \ln\left(\frac{M_\mathrm{join}}{M_\mathrm{piv}}\right) + c_i \ln\left(\frac{M_\mathrm{join}}{M_\mathrm{piv}}\right)^2 = \nonumber\\
    &a_{i+1} + b_{i+1} \ln\left(\frac{M_\mathrm{join}}{M_\mathrm{piv}}\right) + c_{i+1} \ln\left(\frac{M_\mathrm{join}}{M_\mathrm{piv}}\right)^2
\end{align}
and its derivative
\begin{equation}
\label{eq:constraintder}
    b_i + 2c_i \ln\left(\frac{M_\mathrm{join}}{M_\mathrm{piv}}\right) = b_{i+1} + 2c_{i+1} \ln\left(\frac{M_\mathrm{join}}{M_\mathrm{piv}}\right).
\end{equation}
In addition, we require 
\begin{equation}
    c_i<0
\end{equation}
to encapsulate our intuition that the mass function becomes steeper for increasing mass.

We define segments with constant length in log-mass with four segments per decade in mass. We define the fourth segment as the pivot segment for which we define the overall amplitude parameter $a_4$ as well as $b_4$ and $c_4$. With these parameters set, the only free parameters left are the $c_i$ for all other segments; the remaining $a_i$ and $b_i$ are set according to the constraints from Eqs.~\ref{eq:constraintamp} and~\ref{eq:constraintder}. We define $M_\mathrm{piv}=10^{14}M_\odot/h$ to help de-correlate the parameters $a_4$, $b_4$, and $c_4$.

To ensure that the mass function keeps its approximate power-law behavior when extrapolated to even lower masses (outside our analysis domain), we define a $0$-th segment for which we set $c_0=0$ and where $a_0$ and $b_0$ are constrained from the first segment in the usual way. The $0$-th segment spans the mass range $10^{13}<M/(M_\odot/h)<10^{13.1}$ and all subsequent segments span $0.25$~dex in mass as discussed.

In practice, for each redshift, we only fit for the $c_i$ that correspond to segments for which the binned halo catalogs do not vanish. All $c_i$ above the highest halo mass are set to the last $c_i$ where binned halo catalogs exist, meaning that the curvature of the log-mass function remains constant toward even larger masses.


\subsubsection{Likelihood Function and Halo Mass Function Fits}

The input data provided by the simulations for each redshift are the binned halo catalog $\vec N_{\mathrm{sim}}$ and a covariance matrix $\vec\Sigma_\mathrm{sim}$ (see Section~\ref{sec:halo_catalogs}). As the mass bins are rather coarse, we do not assume the mass function to be constant within a bin, and we explicitly model the number of halos in a given bin $i$ as
\begin{equation}
N_{\mathrm{model},\,i} = L_\mathrm{box}^3 \int_{M_{\mathrm{min,}\,i}}^{M_{\mathrm{max,}\,i}}d\ln M \frac{dn}{d\ln M}
\end{equation}
with $dn/d\ln M$ from Eq.~\ref{eq:piecewiseHMF}. We define the log-likelihood function
\begin{equation}
\label{eq:halo_cat_like}
\begin{split}
\vec D &\equiv \vec N_{\mathrm{sim}} - \vec N_\mathrm{model} \\
\ln \mathcal L_\mathrm{sim} &= -\frac12 \vec D^T \vec\Sigma_\mathrm{sim}^{-1} \vec D + \mathrm{const.}
\end{split}
\end{equation}

We regularize the variance in the fit parameters $c_i$ by applying a Gaussian likelihood on the variance between each pair of consecutive $c_i$
\begin{equation}
    \ln \mathcal L_\mathrm{var} = -(i_\mathrm{max}-1)\ln\lambda + \sum_{i=1}^{i_\mathrm{max}-1} -\frac12 \left(\frac{c_i-c_{i+1}}\lambda\right)^2
\end{equation}
up to a constant. The standard deviation $\lambda$ is a free parameter of the model. The role of $\ln \mathcal L_\mathrm{var}$ is to constrain $\lambda$ and the $c_i$ by striking a balance between a good fit (which requires the $c_i$ to differ and thus leads to a non-continuous second derivative of the log-mass function) and a second derivative of $\ln(dn/d\ln M)$ that is as little jumpy as possible.

Finally, for each redshift, we constrain the parameters $a_4$, $b_4$, $c_i$, and $\lambda$ for each design model by sampling the total log-likelihood
\begin{equation}
    \ln \mathcal L = \ln \mathcal L_\mathrm{sim} + \ln \mathcal L_\mathrm{var} + \mathrm{const.}
\end{equation}

The residuals of the best-fit mass functions obtained this way are consistent with the statistical noise in the halo catalogs. The distribution of reduced $\chi^2$ across all models and redshifts is close to being centered at unity.

\begin{figure}
    \includegraphics[width=\columnwidth]{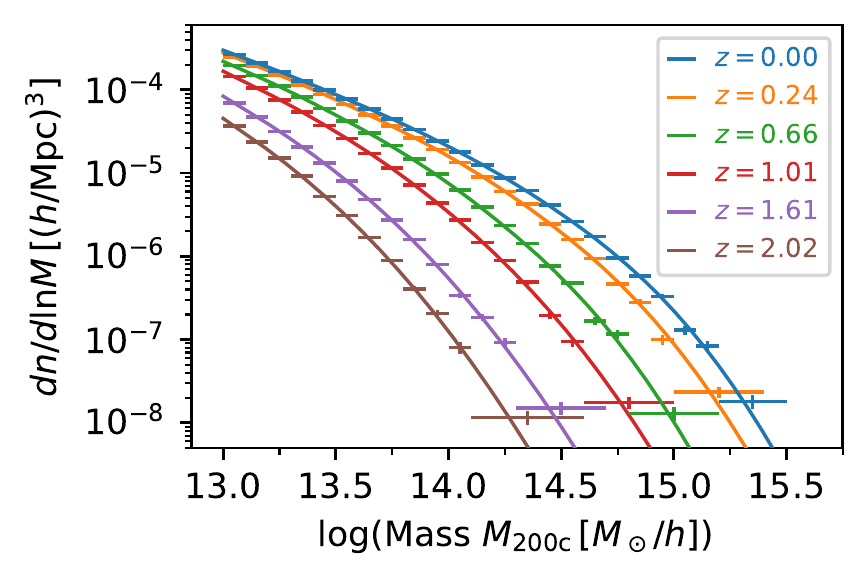}
    \caption{Binned halo catalog and mass function fits for a sample design model (M042). Vertical bars show the Poisson noise on the halo catalogs. We omit two redshifts for the sake of readability.}
    \label{fig:HMFdetail}
\end{figure}

\begin{figure}
    \includegraphics[width=\columnwidth]{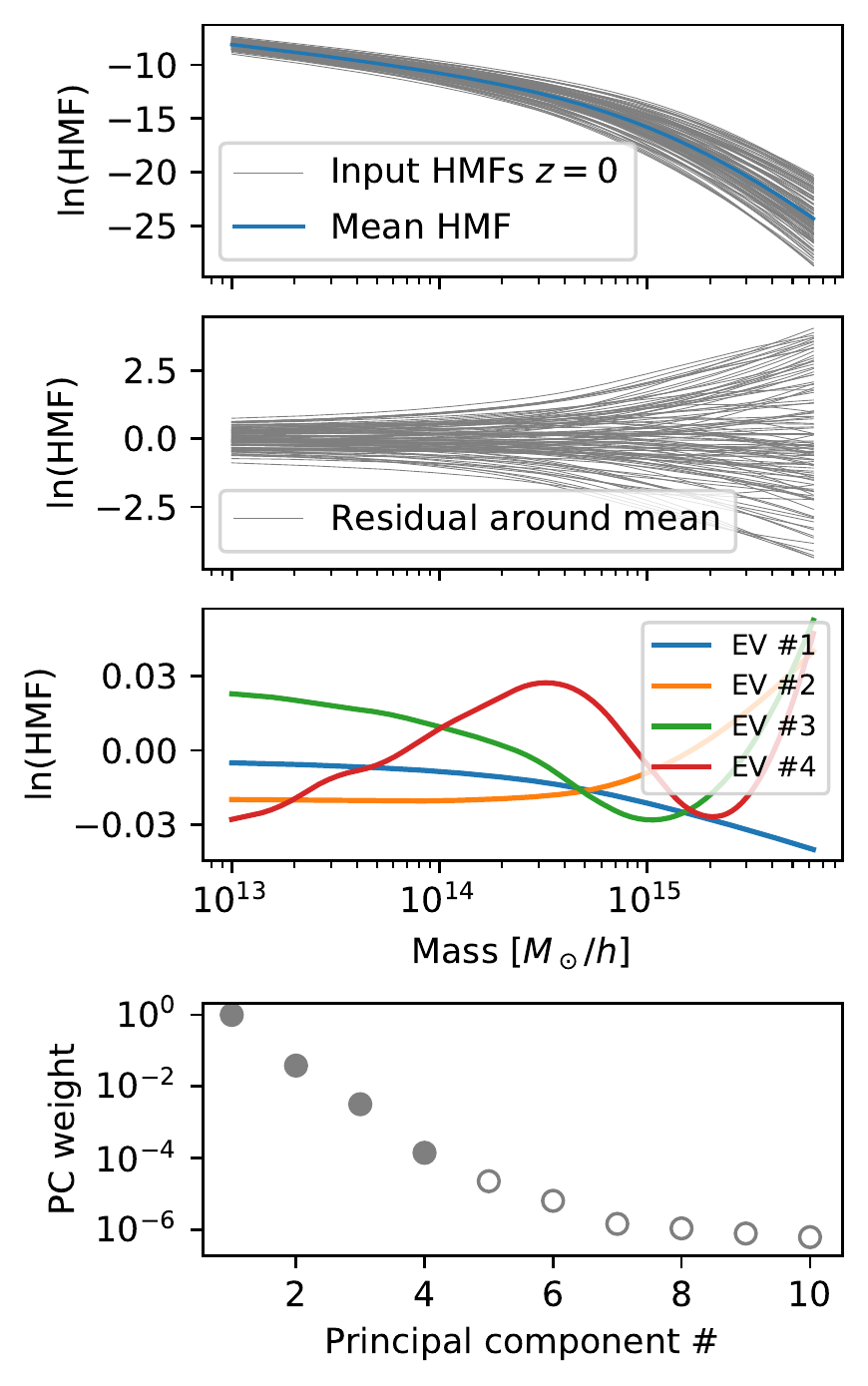}
    \caption{Example of the PCA for the $z=0$ snapshot; this is repeated for each snapshot independently. \emph{Top panel}: Best-fit mass function for each of the 111 design cosmologies along with the mean model. \emph{Second panel:} Fractional residual of each model with respect to the mean. The PCA is performed on these residuals. \emph{Third panel:} Eigenvectors of the first four components. We re-fit this set of basis functions to each model in Section~\ref{sec:refit}. \emph{Bottom panel:} Eigenvalues of the first 10 PCs. We keep the first four PCs, corresponding to the filled circles.}
    \label{fig:PC_z0}
\end{figure}

In Figure~\ref{fig:HMFdetail}, we show the binned halo catalogs and the mass function fits for model M042 for illustration purposes. The range of best-fit mass functions for all models at $z=0$ is shown in the top panel of Figure~\ref{fig:PC_z0}.

\subsection{Basis Functions}
\label{sec:HMFbasis}

From the previous subsection, we now have 111 best-fit mass function models for each snapshot. To compress that information, we perform a PCA. We prepare the best-fit mass function models by taking their logarithm to reduce the dynamic range and by subtracting the mean log-model (see second panel of Figure~\ref{fig:PC_z0}). Because of the exponential high-mass cutoff of the mass function, it is clear that the best-fit mass functions diverge in the high-mass regime. As we do not want this divergence to dominate the PC decomposition, we limit the best-fit mass functions to the mass range covered by the binned halo catalogs. After carrying out some sensitivity tests, we decided to keep the first four principal components and their associated four eigenvectors in mass function space. With four components, the error introduced by the PCA is about 1\% -- although this value is not directly relevant as we re-compute all PC weights in the next subsection. The eigenvectors and eigenvalues for $z=0$ are shown in the two bottom panels of Figure~\ref{fig:PC_z0}.


\subsection{(Re-)Computing the Principal Component Weights}
\label{sec:refit}

The PCA performed in the previous subsection provides, for each snapshot, four eigenvectors (or basis functions) and the PC weights associated with each design model. Note that these PC weights are point-estimates with no uncertainties, because we performed the PCA on the (zero-uncertainty) best-fit mass function models. To propagate the statistical noise in the halo catalogs onto the PC weights, we now (re-)fit the mass function, parametrized by the mean mass function and the four basis functions from the previous subsection, to the halo catalogs. As we are explicitly interested in the uncertainty on the PC weights, we perform a Markov Chain Monte Carlo (MCMC) analysis for each halo catalog using the likelihood function defined in Eq.~\ref{eq:halo_cat_like}.

As a result, the input mass function for every input cosmology and redshift is now described by a four-dimensional posterior parameter distribution. All posterior distributions (8 snapshots $\times$ 111 models) are well-approximated by multi-variate Gaussian distributions and we thus extract the parameter means and covariances for all models.

\begin{figure}
    \includegraphics[width=\columnwidth]{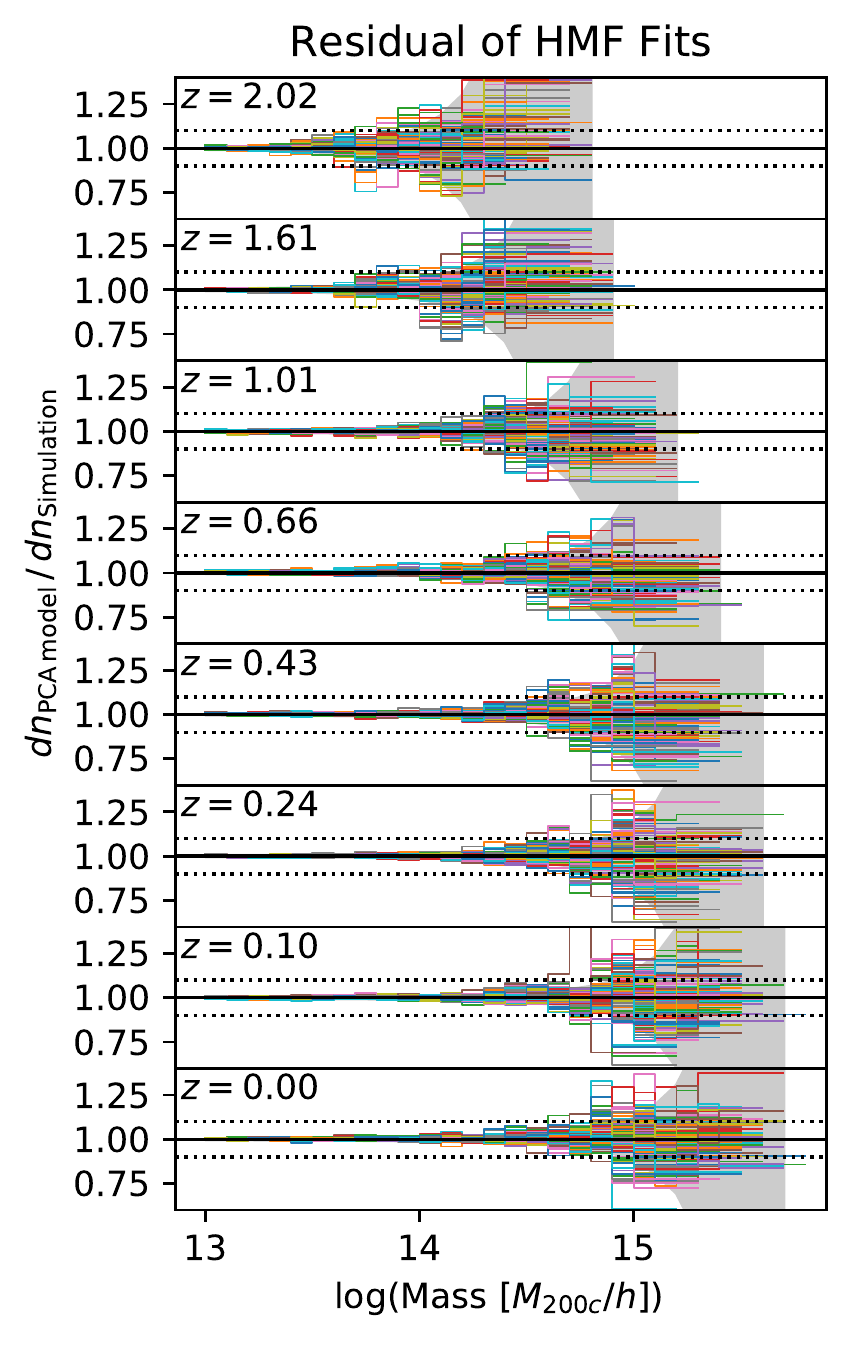}
    \caption{Residual between the halo mass function fits for simulation boxes M001--M111 and their underlying halo catalogs. These mass functions are the input for the actual emulator. The grey shaded bands show the median shot noise across all models as an estimate for the typical statistical noise in the halo catalogs (for visualization purposes only). Dotted lines show $\pm10\%$ to guide the eye. The residuals are consistent with statistical noise in the halo catalogs.}
    \label{fig:inputHMF}
\end{figure}

We validate the description of the mass functions with these four parameters by studying the residuals between each mean mass function parametrized in this way and its underlying halo catalog. The residuals are shown in Figure~\ref{fig:inputHMF} which suggests that they are consistent with statistical noise in the halo catalogs. For each redshift, the distribution of residuals across all models is consistent with a reduced $\chi^2$ of unity.

\subsection{Gaussian Process Regression}
In this section, we briefly review the basics of Gaussian Process (GP) regression and set the appropriate notation. We then apply this method and build an interpolation scheme for the PC weights obtained above.


\subsubsection{Basics of Gaussian Process Regression}
\label{sec:GPintro}
Gaussian Process regression assumes a collection of data values $\vec f$ at locations $\vec X$ that are drawn from a joint Gaussian distribution $\vec f = \mathcal N(\vec 0,\vec K(\vec X))$, with a correlation matrix $\vec K$. Without loss of generality, we have subtracted the mean such that the GP is centered at $\vec 0$. Following, e.g., \cite{Rasmussen2006gpml.book.....R}, the GP can be conditioned on training values $\vec f$ at inputs $\vec X$ with measurement errors $\vec\sigma$ such that it predicts function values $\vec f_\star$ at new locations $\vec X_\star$:
\begin{equation}
\label{eq:GP}
\begin{split}
    &\vec K \equiv \vec K(\vec X,\vec X)\\
    &\vec K_\star \equiv \vec K(\vec X_\star,\vec X)\\
    &\vec K_{\star\star} \equiv \vec K(\vec X_\star,\vec X_\star)\\
    &\vec f_\star|\vec X_\star, \vec X, \vec f \sim \\
    & \mathcal N\Bigl(\vec K_\star \left[\vec K + \vec \sigma \right]^{-1} \vec f, \,
    \vec K_{\star\star} - \vec K_\star\left[\vec K + \vec \sigma \right]^{-1}\vec K_\star\Bigr).
\end{split}
\end{equation}

The correlation matrix $\vec K$ is constructed using the correlation function $k$ between two inputs $\vec x$ and $\vec x^\prime$. A popular choice is the squared exponential kernel, which we adopt in a slightly modified form
\begin{equation}
\label{eq:corr_func_1}
    k(\vec x, \vec x^\prime, \lambda, \vec \rho) = \lambda^{-1} \prod_{i=1}^\Ndimx \rho_i^{4(x_i-x_i^\prime)^2},
\end{equation}
where \Ndimx\ is the dimensionality of $\vec x$ and
with hyperparameters $\lambda$ and $\vec \rho$, setting the precision of the GP and the spatial correlation length between inputs, respectively. For example, a GP with a large precision $\lambda$ varies little along spatial directions, resulting in very smooth functions. As $\rho_i$ approaches unity (from below), the correlation length along direction $i$ becomes infinite and we call the dimension $i$ ``inactive''; conversely, for $\rho_i\ll1$, the spatial correlation length becomes very short. The remaining task is thus to set the hyperparameters in an informed way; we will discuss this in Section~\ref{sec:hyperparams}.

The above summary describes a GP regression scheme for a scalar function $f$ that depends on a multi-dimensional argument $\vec x$. It is straightforward to extend the formalism to vector functions while taking the possible correlations between the function vector elements into account. For \Ndimf-dimensional data, one constructs \Ndimf\ correlation matrices $\vec K$, where each correlation matrix $\vec K_n$ has an independent set of hyperparameters $\lambda_n$ and $\vec\rho_n$. The measurement errors on the input parameters are now stored in a covariance matrix with $\Ndimf \times \Ndesign$ entries on a side, where \Ndesign\ is the number of input data points. With this formalism, we account for correlated input parameters but assume that the hyperparameters for each input are independent. We will employ such a multi-dimensional GP regression scheme for our emulator.


\subsubsection{GP Input Parameters}
We prepare the inputs for the GP regression as follows. The cosmological parameters of the design models\footnote{We choose the $w_0$-$w_b$ parametrization of the dark energy equation of state (see Eq.~\ref{eq:wb}).} are normalized to the range $[0,1]$. The mean PC weights from Section~\ref{sec:refit} are standardized to have zero mean and unit variance. The associated covariance matrices are rescaled accordingly.

Thus, we have $\Ndimx=8$ function arguments (the cosmological parameters), $\Ndimf=4$ dimensional function values (the PC weights), and $\Ndesign=111$ input data points.


\subsubsection{Hyperparameter Optimization}
\label{sec:hyperparams}
The parameters $\lambda_1,\dots,\lambda_\Ndimf$ and $\vec\rho_1,\dots,\vec\rho_\Ndimf$ are free parameters of the GP model that need to be specified. The marginal likelihood of the GP can be interpreted as a likelihood function of the inputs given the model hyperparameters:
\begin{equation} \label{eq:marginal_like}
\begin{split}
        \ln\mathcal L(\vec f|\vec X) =& -\frac12 \vec f^T \left(\vec K + \vec \Sigma\right)^{-1} \vec f \\
        &- \frac12 \ln\bigl[\det(\vec K + \vec \Sigma)\bigr] + \mathrm{const.}
\end{split}
\end{equation}
Following \cite{Habib2007PhRvD..76h3503H}, we apply priors $\pi$ on the hyperparameters
\begin{align}
    0.1 &< \lambda_i < 2.5, \nonumber \\
    \pi(\lambda_i) &\propto \lambda_i^4 \exp(-5\lambda_i), \, &i=1,\dots,\Ndimf \label{eq:prior_a}\\
    0.4 &< \rho_{ij} < 1, \nonumber \\
    \pi(\rho_{ij}) &\propto (1-\rho_{ij})^{-0.9}, \, &i=1,\dots,\Ndimf, \nonumber\\
    &&j=1,\dots,\Ndimx. \label{eq:prior_b}
\end{align}
Because we standardized the GP input parameters in the previous subsection, we expect the variance $\vec\lambda$ to be close to unity and choose a prior with mean of $1$ and standard deviation of $0.45$. Similarly, as the design parameters are normalized to $[0,1]$, the adopted prior on $\vec \rho$ encodes our expectation of very smooth functions (and few active dimensions); we have substantial prior mass near $1$, and $\mathrm{Pr}(\rho_{ij}<0.98)\approx1/3$.

The final step is to set the hyperparameters such that they maximize the likelihood in Eq.~\ref{eq:marginal_like}. This represents an inference problem with $\Ndimf+\Ndimf\times\Ndimx=36$ free parameters. For each snapshot, we run an MCMC analysis and set the hyperparameters to the mean recovered values from the MCMC.


\subsection{Emulator Output and Uncertainty}
\label{sec:emu_out}
For a given requested cosmology, the emulator uses the GP regression scheme to estimate the mean PC weights and their covariance matrix according to Eq.~\ref{eq:GP}. As a first consistency check, we confirm that the residual between the four-dimensional PC weights that we obtained by fitting to the halo catalogs of the design models in Section~\ref{sec:refit} are statistically consistent with the PC weights that the GP returns at those locations. This test confirms that the emulator prediction indeed goes through the design data points.

As a next step, we compute the consistency between the PC weights obtained in Section~\ref{sec:refit} for the test models, which were not used for the construction of the emulator, and the prediction by the GP. For three out the eight snapshots, the reduced $\chi^2$ across the five test models is slightly smaller than unity, indicating that the emulator is making statistically accurate predictions. For the remaining five snapshots, the reduced $\chi^2$ vary between $1.3$ and $2.4$. We multiply the GP output covariance matrix with fixed factors such that the reduced $\chi^2$ is forced to unity. This slightly degrades the emulator precision at the benefit of having statistically consistent residuals with the test models. The emulator is now fully specified.

\begin{figure}
    \includegraphics[width=\columnwidth]{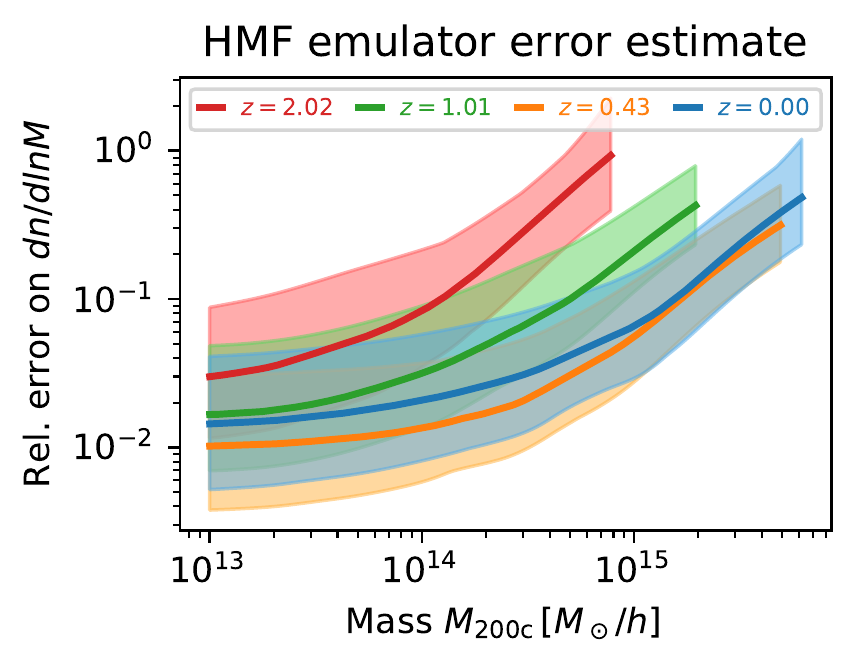}
    \caption{Emulator error estimates.
    The precision also varies as a function of the cosmological parameters (proximity to design models reduces the error). We compute emulator predictions at 1000 random locations within the design hypercube and plot the median emulator error as solid lines and the full range of errors as colored bands (we only show four redshifts for improved readability). The typical precision at $10^{14}\,M_\odot/h$ varies between 1\% and 10\% depending on redshift.}
    \label{fig:HMF_rel_err}
\end{figure}

To obtain predictions for the mass function, the PC weights are multiplied with the mass function basis functions, the mean log-mass function is added, and upon exponentiation we obtain the emulated mass function (reverse process of Section~\ref{sec:HMFbasis}). Note that the mass function computed in this way corresponds to the mean predicted PC weights, and thus contains no uncertainty. To provide an error estimate, we repeat the computation of the mass function, but instead of taking the mean predicted PC weights, we draw realizations from the multivariate probability distribution that the GP provides, accounting for the additional factors discussed just above. We compute the error on the emulator prediction by taking the variance of such a set of mass function predictions from the GP covariance. Note that the error estimates vary with  mass and redshift as they account for the noise in the input halo catalogs; they also vary as a function of the location in cosmology space according to the proximity to design locations (see Figure~\ref{fig:HMF_rel_err}).

Note that in principle there is another contribution to the emulator uncertainty due to the uncertainties in the GP hyperparameters. However, across all redshifts, this additional source of noise is about an order of magnitude smaller than the errors discussed above and we therefore neglect it.


\subsection{Redshift Evolution of the Halo Mass Function}
Our emulator provides the mass function for the requested cosmology for 8 discrete redshifts between 0 and 2. To obtain the mass function at intermediate redshifts, we recommend linearly interpolating the logarithm of the mass function $\ln[dn(M,z)/d\ln M]$.


\subsection{Emulator Performance}
\label{sec:emu_validation}
We verify the emulator performance in two ways: hold-out tests and tests against additional simulations for cosmologies that were not used in the original construction of the emulator.


\begin{figure}
    \includegraphics[width=\columnwidth]{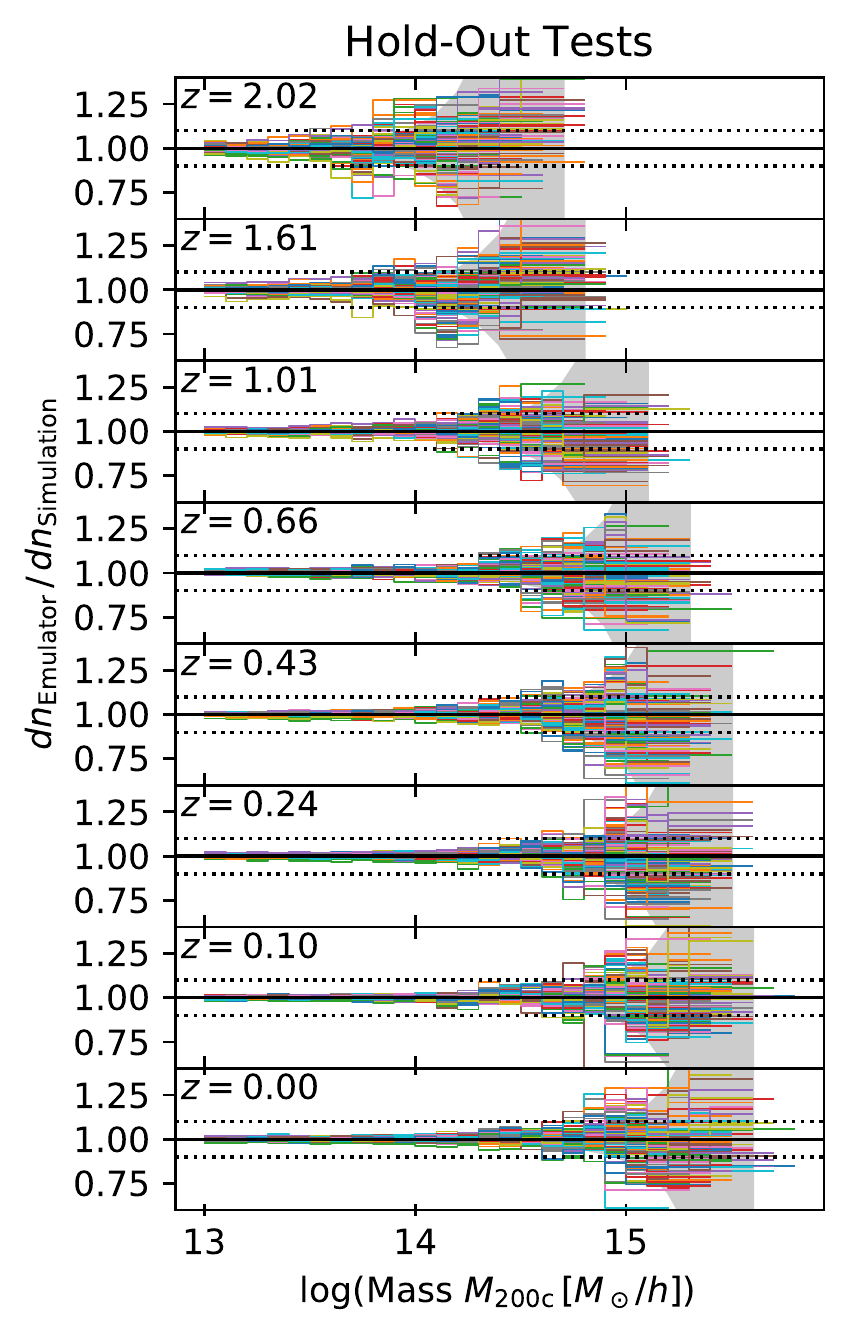}
    \caption{Emulator validation with the hold-out test: The emulator is constructed using all but one model, and the residual of the mass function is plotted for that model. The procedure is repeated for each of the 111 design models. The grey shaded areas show the median shot noise across all models as an estimate for the typical statistical noise in the halo catalogs (for visualization purposes only). Dotted lines show $\pm10\%$ to guide the eye. }
    \label{fig:leaveoneout}
\end{figure}

\subsubsection{Hold-Out Tests}

Hold-out tests are performed by constructing the emulator using all but the $i$-th design model, and this is repeated for each design model $i$. The hyperparameters of the GP are kept fixed to the values obtained for the full design in Section~\ref{sec:hyperparams}. Intuitively, the hold-out tests allow us to assess whether the function to be emulated is smooth enough so that $N_\mathrm{design}-1$ input cosmologies are sufficient for emulation purposes.

The result of the test is shown in Figure~\ref{fig:leaveoneout}. The mass functions are correctly predicted at the few-percent level. At high mass, as expected, this test is limited by the statistical noise in the halo catalogs. Note that the hold-out test is powerful, because we are effectively testing the impact of a ``hole'' in the design, and the full emulator has no such holes. We conclude that our emulator passes this test and that the claim of percent-level mass function predictions made above in Section~\ref{sec:emu_out} and Figure~\ref{fig:HMF_rel_err} are justified.


\subsubsection{Verification With Additional Simulations}
\label{sec:validate_test}

An alternative verification scheme consists of carrying out spot checks by comparing the emulator predictions with halo catalogs obtained for additional cosmologies that were not used to construct the emulator; this avoids the ``sampling hole'' problem (potentially most troublesome near the hypercube boundaries) with hold-outs, at the obvious cost of a more systematic coverage of parameter space.

\begin{figure}
    \includegraphics[width=\columnwidth]{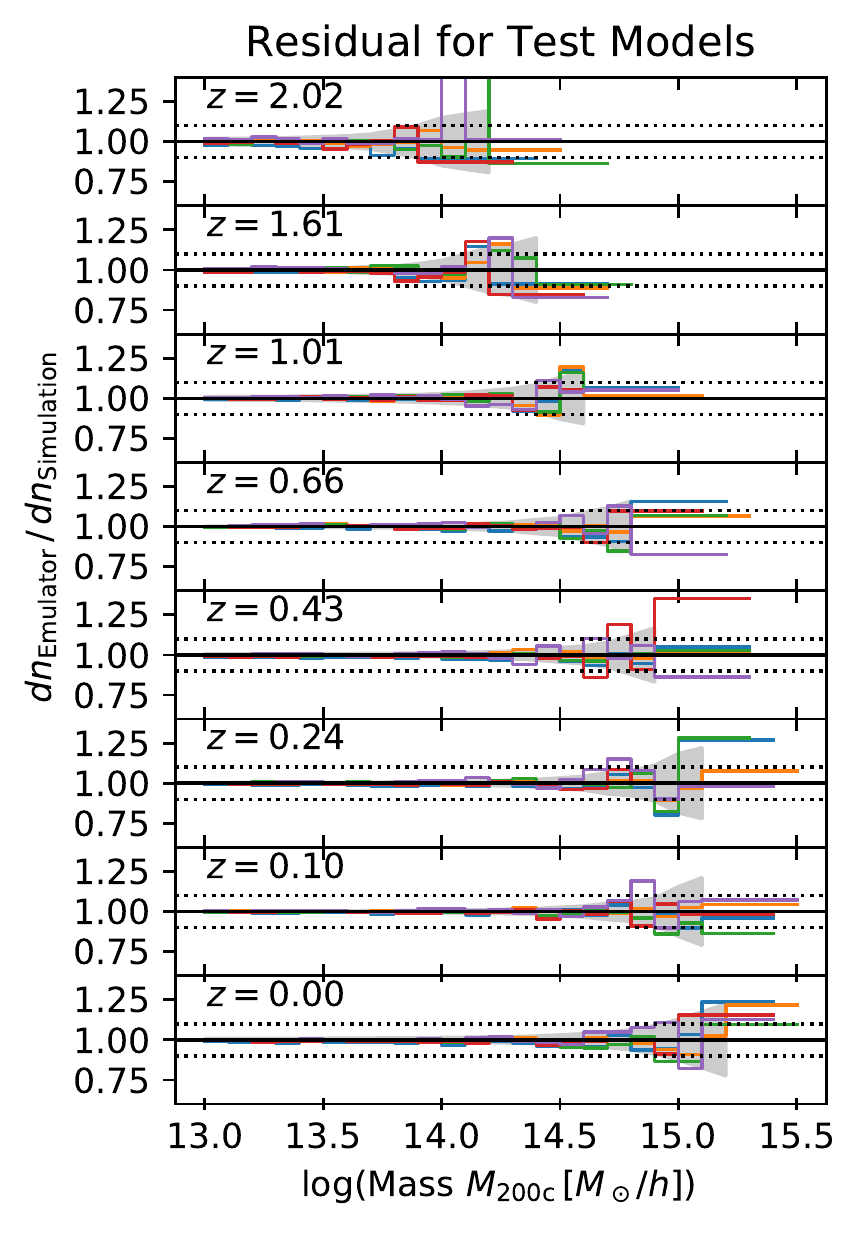}
    \caption{Emulator verification: We compare the emulator prediction with the mass function from additional simulations with cosmologies that were not used in the construction of the emulator. The grey shaded areas show the median shot noise across all models as an estimate for the typical statistical noise in the halo catalogs (for visualization purposes only). Dotted lines show $\pm10\%$ to guide the eye.}
    \label{fig:additional}
\end{figure}

In Figure~\ref{fig:additional}, we compare the emulator prediction with halo catalogs from our five additional cosmologies (M000, T001--T004). Here again, we observe percent-level agreement. By construction, the residuals are within the combined emulator error and the statistical noise in the halo catalogs (see Section~\ref{sec:emu_out}). We thus conclude the verification of the emulator.

\subsection{Concluding Remarks}

We conclude this section with a discussion of the approach adopted for the construction of the emulator. The initial mass function fits described in Section~\ref{sec:HMFfit} are only used to construct a (close to) optimal set of basis functions in the subsequent Section~\ref{sec:HMFbasis}. We then keep four basis functions and obtain four-dimensional parameter sets by fitting these basis functions to the halo catalogs in Section~\ref{sec:refit}. In principle, any other set of basis functions would work, too. One could, for example, directly use the binned halo catalogs as inputs to the PCA. This would, however, lead to non-continuous mass function predictions and higher overall noise levels which is why we did not follow this approach.

In yet another possible analysis setup, we used the universal mass function functional form (described in detail in Section~\ref{sec:universality}) to obtain smooth mass function fits for each model. However, we found that this parametrized mass function is not able to accurately capture the details of the behavior at high mass. The inaccuracy is very subtle: for a given model, the residuals and reduced $\chi^2$ are completely acceptable. However, when measuring the residuals for all 111 models, we noticed that the last one or two bins in the halo catalogs almost all scattered high compared to their respective best-fit mass function model. The distribution of reduced $\chi^2$ for all models indeed suggested a slightly bad fit.

By adopting the approach of using a constrained piecewise polynomial model instead of raw binned halo counts or the universal fitting function approach, we strike a balance between being purely data-driven and incorporating physical intuition about the mass function. In summary, our mass function emulator is built upon the assumptions that the (logarithm of the) mass function can be described by a second-order polynomial in log-mass whose second derivative varies on scales of about 0.25 dex in mass, and that the variation of the (re-fitted) PC weights with cosmology can be captured by a multi-variate Gaussian distribution. The residuals with respect to the design halo catalogs were shown in Figure~\ref{fig:inputHMF}.


\section{Discussion of Existing Halo Mass Function Fits and Emulators}
\label{sec:literatureHMF}

\begin{figure}
    \includegraphics[width=\columnwidth]{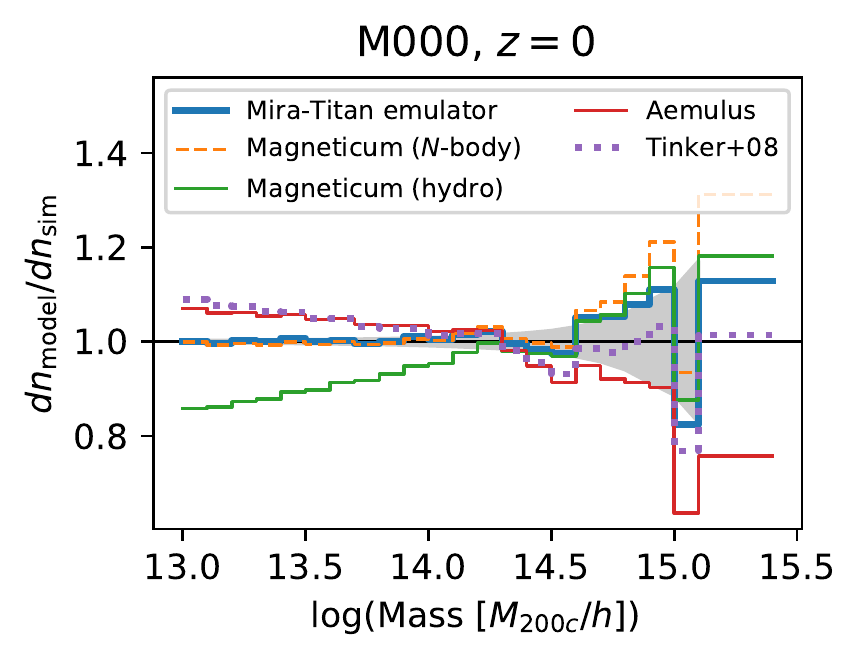}
    \caption{Comparison of different halo mass function predictions with the simulation for the $\Lambda$CDM test model M000. The grey shaded region represents the shot noise in the simulation halo catalog. The fit to the Magneticum simulations and the fit by \cite{Tinker2008ApJ...688..709T} both rely on the universal approach and are calibrated to fiducial $\Lambda$CDM cosmologies. The Magneticum $N$-body fit agrees very well with our simulation; its hydrodynamic counterpart predicts less halos below about $10^{14.2}M_\odot/h$ as shown in \cite{Bocquet2016MNRAS.456.2361B}. At masses below $10^{14}M_\odot/h$, the \cite{Tinker2008ApJ...688..709T} fit and the \Aemulus~emulator overpredict the number of halos because of non-preclusion of halo overlap. The prediction from our emulator agrees well with the simulation, as also shown in Figure~\ref{fig:additional}.}
    \label{fig:literatureHMF}
\end{figure}

We discuss our emulator in the context of existing halo mass function fits and emulators. We consider the simulation for our $\Lambda$CDM (and massless-neutrino) cosmology M000, which was not used for the construction of our emulator, and confront it with different predictions from the literature. We perform this comparison at redshift $z=0$ and show the results in Figure~\ref{fig:literatureHMF}. As discussed above, our emulator provides an accurate prediction for the mass function at the M000 cosmology.

A popular set of fitting functions that covers a range of SO mass definitions is described in \cite{Tinker2008ApJ...688..709T}; we compare that prediction with our simulation in Figure~\ref{fig:literatureHMF}.\footnote{We verify our implementation of the fitting function against the \textsc{HMFcalc} tool and find agreement better than 0.4\% \citep[\url{http://hmf.icrar.org/};][]{Murray2013A&C.....3...23M}.} Above $M\gtrsim10^{14}M_\odot/h$, this fit provides a very good description of our simulation. At lower mass, however, the \cite{Tinker2008ApJ...688..709T} fit overestimates the number of halos by up to about 10\%. This behavior is expected since we extract SO masses for FOF halos, while \cite{Tinker2008ApJ...688..709T} use an SO finder that allows for overlapping subhalos which can boost the number of objects \citep[see the discussion in, e.g., ][]{Kravtsov2004ApJ...609...35K, Tinker2008ApJ...688..709T, Bocquet2016MNRAS.456.2361B}. We compare to an alternative fitting function presented in \cite{Bocquet2016MNRAS.456.2361B} that was calibrated to the Magneticum Pathfinder simulation suite (Dolag et al., in prep.).\footnote{\url{http://www.magneticum.org/}} Importantly, in this work SO masses were extracted for FOF halos defined by $b=0.16$ corresponding to almost the same halo definition as we use (SO masses for $b=0.168$ halos). Indeed, the mass function for the Magneticum $N$-body simulations agrees very well with our M000 model. The mass function for the Magneticum hydrodynamic simulations yields significantly less low-mass clusters and groups than their $N$-body counterparts or our simulation. This is expected, since hydrodynamic feedback effects drive baryons from the halo center, which, for a given halo, leads to a lower total enclosed mass -- this is turn leads to a lower halo abundance at fixed mass \citep[e.g.,][see also the discussion in Section~\ref{sec:summary}]{Henson2017MNRAS.465.3361H, Springel2018MNRAS.475..676S}. In conclusion, after having shown that our emulator passes the verification tests against our own simulations in the previous section, we have now shown that, for our $\Lambda$CDM model, our simulation and emulator also agree with two example fitting functions from the literature.\footnote{In Section~\ref{sec:universality}, we show that such a verification is not possible for the other cosmologies of the \MiraTitanUniverse\ suite as the universal approach is not accurate enough to correctly capture models with massive neutrinos and dynamical dark energy.}

To date, two emulators for the halo mass function have been presented: the \DarkEmulator\ \citep{Nishimichi2019ApJ...884...29N} and \Aemulus, which is publicly available\footnote{\url{https://github.com/AemulusProject}} \citep{McClintock2019ApJ...872...53M}. Both of these are only valid for massless neutrinos and a constant dark energy equation of state while our emulator additionally covers massive neutrinos in the parameter range $0\leq\Onuhh\leq0.01$ and dynamical dark energy through the $w_0$-$w_a$ parametrization (see Section~\ref{sec:simulations}). As a result, only our emulator can be used for cosmological analyses that aim at constraining the sum of neutrino masses and/or dynamical dark energy.

The \DarkEmulator\ covers a six-dimensional cosmological parameter space. As opposed to our emulator, it is constrained to $\sumMnu=0$ and a constant dark energy equation of state ($w_a=0$).
Their simulation design contains 101 cosmologies which are sampled from a six-dimensional hypercube. A sliced latin hypercube design is employed to allow for an even sampling of the parameter space (as a reminder, we achieve even sampling through a tesselation-based nested design strategy, see Section~\ref{sec:simulations}). For each model, $1\,\mathrm{Gpc}/h$ and $2\,\mathrm{Gpc}/h$ simulation boxes are run evolving $2048^3$ particles each; only the smaller boxes are used for the halo mass function emulator. In summary, the \DarkEmulator\ framework is similar to ours except for the more restrictive range of cosmologies.

The \Aemulus\ suite of simulations \citep{DeRose2019ApJ...875...69D} covers seven cosmological parameters; compared to our work, they impose $\Onuhh=0$ and $w=\mathrm{const.}$ (equivalent to $w_a=0$) but vary the effective number of relativistic species $N_\mathrm{eff}$. Importantly, their design cosmologies are chosen to sample the parameter space allowed by the union of the cosmological constraints from WMAP9+BAO+SNIa and Planck13+BAO+SNIa data and their emulator cannot produce accurate results outside of this region of the parameter space. In contrast, the \MiraTitanUniverse\ design samples the entire parameter hypercube, as described in Section~\ref{sec:simulations}. Therefore, this hypercube can be adopted as hard priors in any cosmological analysis that uses our emulator, whereas the \Aemulus\ emulator can only be used in analyses where strong priors from CMB, BAO, and SNIa measurements are applied. The 40 \Aemulus\ simulations have similar mass resolution as ours but smaller box sizes of $1.05\,\mathrm{Gpc}/h$, thus providing a less precise calibration of the number (density) of higher-mass halos. As a cross-check, we ran the \Aemulus\ emulator for our M000 cosmology. Note that \Aemulus\ provides SO masses \Mtwom\ (using the \textsc{rockstar} halo finder and thus not the same halo definition as we do); we convert to \Mtwoc\ assuming an NFW profile \citep{Navarro1997ApJ...490..493N} and the concentration--mass relation by \cite{Child2018ApJ...859...55C}.\footnote{We use the Colossus package: \url{https://bitbucket.org/bdiemer/colossus}} As shown in Figure~\ref{fig:literatureHMF} and demonstrated in \cite{McClintock2019ApJ...872...53M}, the prediction from \Aemulus\ is very similar to the prediction by \cite{Tinker2008ApJ...688..709T}: it agrees with our mass function at about $10^{14}M_\odot/h$ and above (within the noise at high mass) and overestimates the halo abundance toward lower masses by up to about 10\%. We attribute the latter effect to different halo definitions, as also discussed in our comparison with \cite{Tinker2008ApJ...688..709T}.

The discrepancy between (some of) the mass function predictions below $\sim 10^{14}M_\odot/h$, which arises because of different halo (center) definitions has two important consequences. The first relates to the fact that cosmological analyses of real data must use simulation inputs in a self-consistent manner, i.e., that the mass definition for which the halo mass function is predicted has to be the same mass definition for which the projected halo mass profiles for weak-lensing measurements are used (in other words, all else being equal, the appropriate mass-observable relation will compensate for changes in the mass definition). The second point is more complicated and has to do with halo identification -- the halo-finding strategy employed in the simulation must correctly take into account how observed clusters are found in practice (see, e.g., \citealt{Garcia2019MNRAS.489.4170G} for additional discussion and an explicit example for optically selected clusters). There must be a consistent pathway to take the halos from a simulated light cone catalog and associate them with single objects that can be found from a simulated observation. (Merging groups and clusters are one example where more care may be needed.) Upcoming and future surveys that push to low cluster masses around $10^{14}M_\odot/h$ and below will need to take this effect into account.


\section{The Limits of Cosmological Universality of the Halo Mass Function}
\label{sec:universality}

Prior to the emergence of cosmological emulators based on numerical simulations, the cluster cosmology community followed an approach based on the assumed universal form for the mass function. As discussed in the Introduction, in this approach, all of the mass function dependence on cosmology and redshift is fully described as a function of the RMS fluctuations $\sigma(M, z)$ in the linear matter power spectrum $P(k,z,\mathrm{cosmology})$. In this section, we use the \MiraTitanUniverse\ simulations to investigate the limits of the assumed \emph{cosmological universality} of the mass function for \Mtwoc.

As a first step, we establish a mass function fit following the universal approach. We choose model M000 (\LCDM\ with massless neutrinos) as our baseline model for which we establish the universal fit; we then extrapolate this result to all other \MiraTitanUniverse\ cosmologies and compare to the simulation results.

The universal parametrization of the halo mass function reads
\begin{equation} \label{eq:HMF}
dn/dM=f(\sigma)\frac{\rho_\mathrm m}M\frac{d\ln \sigma^{-1}}{dM}
\end{equation}
with $\rho_\mathrm m$ designating the matter density. The RMS fluctuation $\sigma$ in the matter density field is related to the linear matter power spectrum $P(k,z)$ via
\begin{equation}
\label{eq:sigma}
 \sigma^2(M,z) = \frac1{2\pi^2} \int P(k,z) \hat W^2(kR) k^2 dk
\end{equation}
with the Fourier transform $\hat W$ of the real-space top-hat window function of radius
\begin{equation} \label{eq:RofM}
R = \left(\frac{3M}{4\pi\rho_\mathrm{m}}\right)^{1/3}.
\end{equation}
A common parametrization of $f(\sigma)$ is
\begin{equation} \label{eq:fsigma}
f(\sigma) = B\left[\left(\frac\sigma e\right)^{-d} + \sigma^{-f}\right]\exp(-g/\sigma^2)
\end{equation}
with fit parameters $d$, $e$, $f$, and $g$ \citep[e.g.,][]{Warren2006ApJ...646..881W, Tinker2008ApJ...688..709T} together with a normalization $B$ set by the condition $\int f(\sigma)d\ln\sigma^{-1}=1$ which is satisfied for
\begin{equation}
    B = 2\left[e^d g^{-d/2} \Gamma(d/2) + g^{-f/2}\Gamma(f/2)\right]^{-1}.
\end{equation}
This model has considerable freedom as it contains an overall amplitude, the slope of the power-law part of the mass function, a relative amplitude of the power-law component, and an exponential cutoff at high mass (low $\sigma$).

For cosmological models with massive neutrinos, the degree of universality can be improved by choosing the appropriate power spectrum $P(k)$ that enters Eq.~\ref{eq:sigma} and matter density $\rho_\mathrm m$ that enters Eq.~\ref{eq:HMF} and \ref{eq:RofM}. Following e.g., \cite{Ichiki2012PhRvD..85f3521I, Costanzi2013JCAP...12..012C, Biswas2019arXiv190110690B} we set the matter density to the density of collapsing matter:
\begin{equation}
    \rho_\mathrm m = \rho_\mathrm{CDM}+\rho_\mathrm{b};
\end{equation}
this is motivated by the fact that neutrinos hardly cluster on halo scales. Following the same reasoning, we use
\begin{equation}
P(k,z) = P_\mathrm{CDM+b}(k,z) \left(\frac{\rho_\mathrm{CDM}(z)+\rho_\mathrm{b}(z)}{\rho_\mathrm{CDM}(z)+\rho_\mathrm{b}(z)+\rho_\nu(z)}\right)^2
\end{equation}
which is the CDM+baryon power spectrum modified to account for the background density evolution of all particle species.

We obtain parameter posterior distributions for the universal parameters by fitting the mass function from Eq.~\ref{eq:HMF} to the M000 halo catalog at $z=0$ using the likelihood function from Eq.~\ref{eq:halo_cat_like}. We then use this set of parameters to predict the mass function for every model M001--M111 and the test models T001--T004.

\begin{figure}
    \includegraphics[width=\columnwidth]{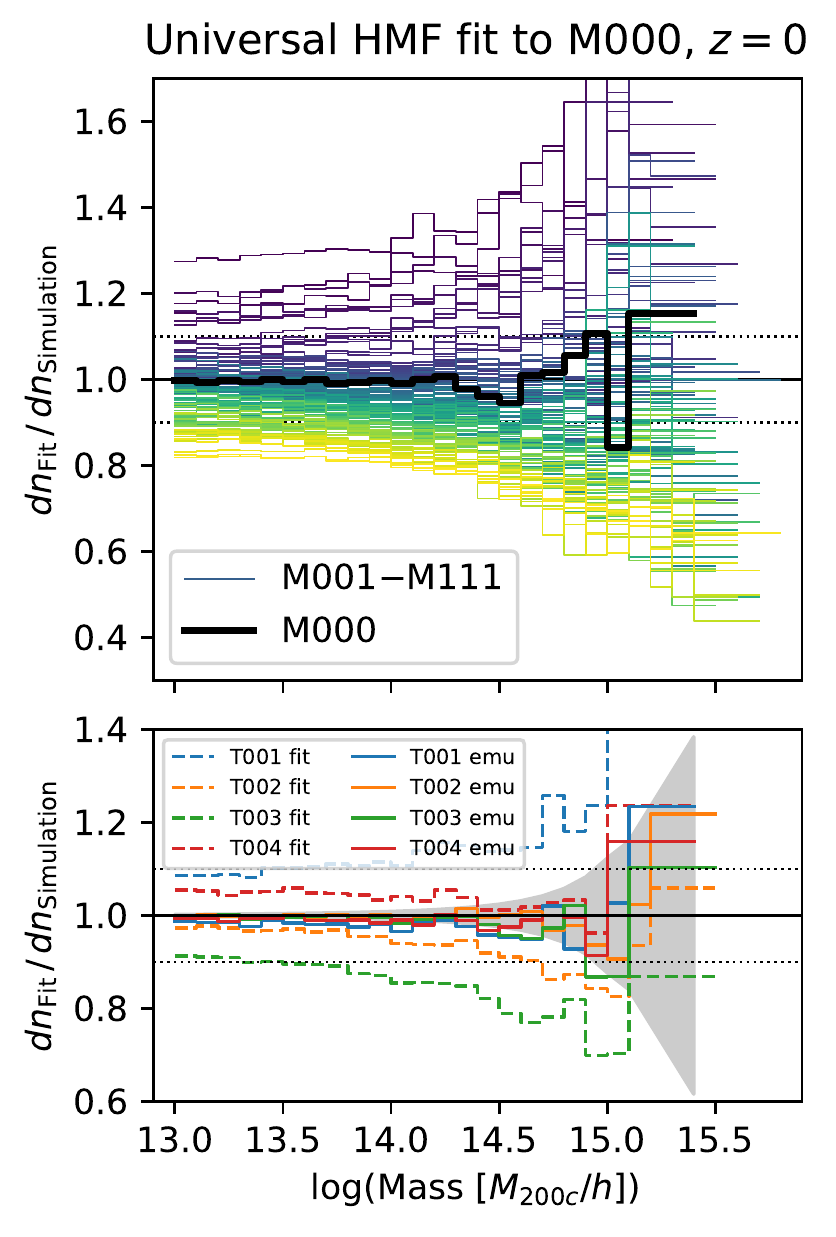}
    \caption{Limits of cosmological universality of the halo mass function: We fit the usual analytical mass function fitting function to the halo catalog from the M000 simulation and compare the residuals between the extrapolation of that fit and all simulations of the \MiraTitanUniverse. In both panels, dotted lines represent $\pm10\%$ errors to guide the eye.
    \emph{Top panel}: The universal mass function fitting function provides a good fit to the data from M000 (thick black line). However, the universal mass function extrapolates poorly to the entire range of cosmologies covered in this work. At $10^{14}M_\odot/h$ for example, residuals range from about $-20\%$ to about $+30\%$.
    \emph{Bottom panel:} Same as top panel, except that we show the residuals for our test simulations T001--T004.
    We also show the emulator residuals which are significantly better. The grey shaded area shows the error due to shot noise in the simulation box.
    }
    \label{fig:universality}
\end{figure}

In the top panel of Figure~\ref{fig:universality}, the thick black line indicates that the universal mass function fit obtained for the M000 model indeed provides a good fit to the M000 halo catalog. However, as we use the set of parameters $d$, $e$, $f$, and $g$ to predict the universal mass function for every model M001--M111, large discrepancies become apparent. Up to masses of about $10^{14}M_\odot/h$, the residuals are approximately constant with mass and range from about $-20\%$ up to about $+30\%$. At higher masses, as we approach the exponential tail of the mass function, the residuals increase significantly.

Note that the degree to which the mass function is universal depends on the specific halo mass definition. For example, the mass function for FOF masses with $b=0.2$ or for SO masses $M_{180\mathrm{m}}$ was shown to feature a more universal behavior for \LCDM\ cosmologies than the mass function for \Mtwoc\ \citep{White2002ApJS..143..241W}. To extend this statement beyond \LCDM, \cite{Bhattacharya2011ApJ...732..122B} calibrated a FOF $b=0.2$ mass function fit using \LCDM\ simulations and then used a suite of 38 simulations carried out for \wCDM\ cosmologies to estimate that the universal approach is still valid at the $\sim10\%$ level. Additional spot checks using a simulation with a \LCDM\ cosmology with massive neutrinos ($\nu\Lambda$CDM) and a simulation with additional dynamical dark energy ($\nu$DDE) were performed by \cite{Biswas2019arXiv190110690B}. They found that the universal mass function approach is able to describe the mass function in their $\nu\Lambda$CDM cosmology to within 5\%, and the mass function in their $\nu$DDE model at the $\sim5\%$ level. In conclusion, we expect that the mass function for \Mtwoc, which we consider in this work, is less universal than for other mass definitions, and FOF masses with $b=0.2$ in particular. Note however, that the formation and evolution of halos is better understood for SO masses that are defined with respect to the critical density, such as \Mtwoc\ or $M_{500\mathrm{c}}$, which justifies using these mass definitions despite the mass function being less universal \citep[see, e.g., the discussion in][]{White2002ApJS..143..241W}. Finally, we note that for \emph{practical} applications, having a sufficiently accurate emulator makes the search for a universal mass function prediction unnecessary. While a universal mass function with a well-characterized uncertainty could be used to expand the cosmological parameter space for which the emulator is valid, the accuracy of the universal mass function outside of the emulator domain of validity would still need to be verified using additional numerical simulations. It remains unclear whether much is gained compared to building a new emulator from such an enhanced simulation campaign.

In the bottom panel of Figure~\ref{fig:universality}, we show the same result as in the upper panel but for the test models T001--T004. As these models were not used for the construction of the emulator, we can directly compare the performance of the universal approach with the emulator. For these models, the universal approach is accurate at the $\sim10\%$ level up to about $10^{14}M_\odot/h$ and worsens somewhat toward higher masses. In contrast, the emulator clearly performs much better as its accuracy stays well within 10\% up to $10^{15}M_\odot/h$.


\section{Summary and Outlook}
\label{sec:summary}

We present, and make publicly available\footnote{This work uses version v0.1.0 \citep{sebastian_bocquet_2020_3895603} of the code available at \url{https://github.com/SebastianBocquet/MiraTitanHMFemulator}. The documentation is hosted at \url{https://miratitanhmfemulator.readthedocs.io/}.}, an emulator for the halo mass function for redshifts $z\leq2$ and for masses $\Mtwoc\geq10^{13}M_\odot/h$ (group and cluster-scale halos) using the newly completed \MiraTitanUniverse\ suite of 111 cosmological $N$-body simulations. The parameter space spans eight cosmological parameters \{\Omhh, \Obhh, \Onuhh, \sig, $h$, $n_s$, \wo, \wa\}.
Our work expands the parameter space covered by existing mass function emulators \citep{McClintock2019ApJ...872...53M, Nishimichi2019ApJ...884...29N} by also including a dynamical dark energy equation of state and the neutrino mass sum. Following our previous philosophy in terms of sampling parameter space, our simulation design covers the entire parameter hypercube, instead of being centered around the currently available cosmological constraints (as employed, e.g., by the \Aemulus\ emulator). Finally, the use of large simulation boxes (2.1~Gpc) yields better statistics for the high-mass tail of the mass function.

We construct the emulator using Gaussian Process regression. In contrast to the mass function emulators in the literature, we do not perform the GP regression on the fit parameters of a universal fitting function, but on a set of principal component weights. The corresponding PC basis functions are obtained in a data-driven way using piecewise polynomial functions. In our approach, we track the noise in the input halo catalogs and the uncertainties involved in the GP regression. We verify the emulator prediction using hold-out tests and tests against additional simulations that were not used for the construction of the emulator.

The accuracy of the emulator is summarized in Figure~\ref{fig:HMF_rel_err}. The accuracy is a function of mass, redshift, and location in cosmological parameter space. For example, at $M=10^{14}M_\odot/h$ and $z<1$, the error is at the percent-level for all cosmologies; at $M=10^{15}M_\odot/h$ it stays below 10\%.

As pointed out in Section~\ref{sec:literatureHMF}, it is important that observational studies use consistent halo definitions for all simulation-based data products that enter the analysis. Furthermore, studies of deep cluster surveys that extend down to $10^{14}M_\odot/h$ or below need to ensure that the halos identified in the simulations match the objects identified in the data. For example, the halo finder may identify two objects (that may partially overlap) when a survey with finite spatial resolution would discover a single, merged cluster. Forward-modeling the cluster abundance as a function of the observable quantities (instead of going through the halo mass function) directly from the simulations may be a promising way to address the issue.

This work does not address the influence of non-gravitational feedback effects on the mass function. This feedback, which is mainly driven by active galactic nuclei, redistributes the material in the cluster center and thereby alters the halo mass profile \citep[e.g.,][]{Henson2017MNRAS.465.3361H, Springel2018MNRAS.475..676S}. The impact on low-mass clusters and groups is more important than on massive clusters which results in the mass function being much less affected in the high-mass regime \citep[e.g.,][]{Cusworth2014MNRAS.439.2485C, Schaller2015MNRAS.451.1247S, Bocquet2016MNRAS.456.2361B}. As a first step, our gravity-only mass function may be modified in post-processing to approximately account for the impact of baryonic effects. Recent work using a set of six hydrodynamic BAHAMAS simulations which include the effects of dynamical dark energy suggests that the impact of baryonic and cosmological effects on the mass function can be separated with an accuracy of $\approx1-2\%$ \citep{Pfeifer2020arXiv200407670P}. It remains to be seen whether such a modified mass function prediction is accurate enough for cosmological analyses of future, high-quality datasets. If not, the mass function prediction will have to be built from large suites of hydrodynamic simulations, which are expected to be available in the near future. In this case, the systematic uncertainty in the hydrodynamic modeling will have to be accounted for and described along with the cosmological parameters in a rigorous statistical analysis framework. The emulator framework presented in this work represents an important step in this direction.


\acknowledgments

We thank Derek Bingham for many discussions and for his contributions to the tessellation-based sampling scheme used here. We also record our debt to Dave Higdon for generously sharing many contributions and sharp insights for more than a decade.
We further thank J\"{o}rg Dietrich and Sebastian Grandis for fruitful discussions about vector-function Gaussian processes and the mass function fits, respectively.
We thank Takahiro Nishimichi and the \DarkEmulator\ team for sharing some of their input mass functions for cross-checking purposes. We thank the referee for a number of thoughtful comments and suggestions that have helped improve the paper.

SB acknowledges the support by the ORIGINS Cluster (funded by the Deutsche Forschungsgemeinschaft (DFG, German Research Foundation) under Germany's Excellence Strategy -- EXC-2094 -- 390783311) and the Ludwig-Maximilians-Universit\"{a}t M\"{u}nchen.
SH, KH, and EL acknowledge support from the U.S. Department of Energy, Office of Science, Office of Advanced Scientific Computing Research and Office of High Energy Physics, Scientific Discovery through Advanced Computing (SciDAC) program under Award Number 231018. 
The work of the authors at Argonne National Laboratory was supported under the U.S. Department of Energy contract DE-AC02-06CH11357. This research used resources of the Argonne Leadership Computing Facility, which is a DOE Office of Science User Facility supported under contract DE-AC02-06CH11357. This research also used resources of the Oak Ridge Leadership Computing Facility, which is a DOE Office of Science User Facility supported under Contract DE-AC05-00OR22725.

\software{
\textsc{astropy} \citep{AstropyCollaboration2018AJ....156..123A},
\textsc{GetDist} \citep{Lewis:2019xzd},
\textsc{matplotlib} \citep{Hunter2007CSE.....9...90H},
\textsc{numpy} \citep{vanderWalt2011CSE....13b..22V},
\textsc{polychord} \citep{Handley2015MNRAS.450L..61H},
\textsc{pyGTC} \citep{Bocquet2016JOSS....1...46B},
\textsc{(py)MultiNest} \citep{Feroz2019OJAp....2E..10F, Buchner2014A&A...564A.125B},
\textsc{scipy} \citep{2020SciPy-NMeth}
}

\appendix

We present all cosmologies of the \MiraTitanUniverse\ and of the additional test models in Table~\ref{tab:cosmo}.

\startlongtable
\begin{deluxetable}{lcccccccc}
\tablecaption{\label{tab:cosmo}Cosmologies in the \MiraTitanUniverse.}
\tablehead{\colhead{Model} & \colhead{$\Omega_\mathrm m h^2$} & \colhead{$\Omega_\mathrm b h^2$}
& \colhead{$\sigma_8$} & \colhead{$h$} & \colhead{$n_s$} & \colhead{$w_0$} & \colhead{$w_a$} & \colhead{$\Omega_\nu h^2$}}
\startdata
\multicolumn{9}{l}{Massless-neutrino design}\\
M001 & 0.1472 & 0.02261 & 0.8778 & 0.6167 & 0.9611 & -0.7000 & 0.67220 & 0.0\\
M002 & 0.1356 & 0.02328 & 0.8556 & 0.7500 & 1.0500 & -1.0330 & 0.91110 & 0.0\\
M003 & 0.1550 & 0.02194 & 0.9000 & 0.7167 & 0.8944 & -1.1000 & -0.28330 & 0.0\\
M004 & 0.1239 & 0.02283 & 0.7889 & 0.5833 & 0.8722 & -1.1670 & 1.15000 & 0.0\\
M005 & 0.1433 & 0.02350 & 0.7667 & 0.8500 & 0.9833 & -1.2330 & -0.04445 & 0.0\\
M006 & 0.1317 & 0.02150 & 0.8333 & 0.5500 & 0.9167 & -0.7667 & 0.19440 & 0.0\\
M007 & 0.1511 & 0.02217 & 0.8111 & 0.8167 & 1.0280 & -0.8333 & -1.00000 & 0.0\\
M008 & 0.1200 & 0.02306 & 0.7000 & 0.6833 & 1.0060 & -0.9000 & 0.43330 & 0.0\\
M009 & 0.1394 & 0.02172 & 0.7444 & 0.6500 & 0.8500 & -0.9667 & -0.76110 & 0.0\\
M010 & 0.1278 & 0.02239 & 0.7222 & 0.7833 & 0.9389 & -1.3000 & -0.52220 & 0.0\\
\tableline
\multicolumn{9}{l}{Main design}\\
M011 & 0.1227 & 0.0220 & 0.7151 & 0.5827 & 0.9357 & -1.0821 & 1.0646 & 0.000345\\
M012 & 0.1241 & 0.0224 & 0.7472 & 0.8315 & 0.8865 & -1.2325 & -0.7646 & 0.001204\\
M013 & 0.1534 & 0.0232 & 0.8098 & 0.7398 & 0.8706 & -1.2993 & 1.2236 & 0.003770\\
M014 & 0.1215 & 0.0215 & 0.8742 & 0.5894 & 1.0151 & -0.7281 & -0.2088 & 0.001752\\
M015 & 0.1250 & 0.0224 & 0.8881 & 0.6840 & 0.8638 & -1.0134 & 0.0415 & 0.002789\\
M016 & 0.1499 & 0.0223 & 0.7959 & 0.6452 & 1.0219 & -1.0139 & 0.9434 & 0.002734\\
M017 & 0.1206 & 0.0215 & 0.7332 & 0.7370 & 1.0377 & -0.9472 & -0.9897 & 0.000168\\
M018 & 0.1544 & 0.0217 & 0.7982 & 0.6489 & 0.9026 & -0.7091 & 0.6409 & 0.006419\\
M019 & 0.1256 & 0.0222 & 0.8547 & 0.8251 & 1.0265 & -0.9813 & -0.3393 & 0.004673\\
M020 & 0.1514 & 0.0225 & 0.7561 & 0.6827 & 0.9913 & -1.0101 & -0.7778 & 0.009777\\
M021 & 0.1472 & 0.0221 & 0.8475 & 0.6583 & 0.9613 & -0.9111 & -1.5470 & 0.000672\\
M022 & 0.1384 & 0.0231 & 0.8328 & 0.8234 & 0.9739 & -0.9312 & 0.5939 & 0.008239\\
M023 & 0.1334 & 0.0225 & 0.7113 & 0.7352 & 0.9851 & -0.8971 & 0.3247 & 0.003733\\
M024 & 0.1508 & 0.0229 & 0.7002 & 0.7935 & 0.8685 & -1.0322 & 1.0220 & 0.003063\\
M025 & 0.1203 & 0.0230 & 0.8773 & 0.6240 & 0.9279 & -0.8282 & -1.5005 & 0.007024\\
M026 & 0.1224 & 0.0222 & 0.7785 & 0.7377 & 0.8618 & -0.7463 & 0.3647 & 0.002082\\
M027 & 0.1229 & 0.0234 & 0.8976 & 0.8222 & 0.9698 & -1.0853 & 0.8683 & 0.002902\\
M028 & 0.1229 & 0.0231 & 0.8257 & 0.6109 & 0.9885 & -0.9311 & 0.8693 & 0.009086\\
M029 & 0.1274 & 0.0228 & 0.8999 & 0.8259 & 0.8505 & -0.7805 & 0.5688 & 0.006588\\
M030 & 0.1404 & 0.0222 & 0.8232 & 0.6852 & 0.8679 & -0.8594 & -0.4637 & 0.008126\\
M031 & 0.1386 & 0.0229 & 0.7693 & 0.6684 & 1.0478 & -1.2670 & 1.2536 & 0.006502\\
M032 & 0.1369 & 0.0215 & 0.8812 & 0.8019 & 1.0005 & -0.7282 & -1.6927 & 0.000905\\
M033 & 0.1286 & 0.0230 & 0.7005 & 0.6752 & 1.0492 & -0.7119 & -0.8184 & 0.007968\\
M034 & 0.1354 & 0.0216 & 0.7018 & 0.5970 & 0.8791 & -0.8252 & -1.1148 & 0.003620\\
M035 & 0.1359 & 0.0228 & 0.8210 & 0.6815 & 0.9872 & -1.1642 & -0.1801 & 0.004440\\
M036 & 0.1390 & 0.0220 & 0.8631 & 0.6477 & 0.8985 & -0.8632 & 0.8285 & 0.001082\\
M037 & 0.1539 & 0.0224 & 0.8529 & 0.5965 & 0.8943 & -1.2542 & 0.8868 & 0.003549\\
M038 & 0.1467 & 0.0227 & 0.7325 & 0.5902 & 0.9562 & -0.8019 & 0.3628 & 0.007077\\
M039 & 0.1209 & 0.0223 & 0.8311 & 0.7327 & 0.9914 & -0.7731 & 0.4896 & 0.001973\\
M040 & 0.1466 & 0.0229 & 0.8044 & 0.8015 & 0.9376 & -0.9561 & -0.0359 & 0.000893\\
M041 & 0.1274 & 0.0218 & 0.7386 & 0.6752 & 0.9707 & -1.2903 & 1.0416 & 0.003045\\
M042 & 0.1244 & 0.0230 & 0.7731 & 0.6159 & 0.8588 & -0.9043 & 0.8095 & 0.009194\\
M043 & 0.1508 & 0.0233 & 0.7130 & 0.8259 & 0.9676 & -1.0551 & 0.3926 & 0.009998\\
M044 & 0.1389 & 0.0224 & 0.8758 & 0.6801 & 0.9976 & -0.8861 & -0.1804 & 0.008018\\
M045 & 0.1401 & 0.0228 & 0.7167 & 0.6734 & 0.9182 & -1.2402 & 1.2155 & 0.006610\\
M046 & 0.1381 & 0.0224 & 0.7349 & 0.8277 & 1.0202 & -1.1052 & -1.0533 & 0.006433\\
M047 & 0.1411 & 0.0216 & 0.7770 & 0.7939 & 0.9315 & -0.8042 & 0.7010 & 0.003075\\
M048 & 0.1374 & 0.0226 & 0.7683 & 0.6865 & 0.8576 & -1.1374 & -0.5106 & 0.004548\\
M049 & 0.1339 & 0.0217 & 0.7544 & 0.5920 & 1.0088 & -0.8520 & -0.7438 & 0.003512\\
M050 & 0.1337 & 0.0233 & 0.8092 & 0.7309 & 0.9389 & -0.7230 & 0.6920 & 0.005539\\
M051 & 0.1514 & 0.0222 & 0.7433 & 0.6502 & 0.8922 & -0.9871 & 0.8803 & 0.002842\\
M052 & 0.1483 & 0.0230 & 0.7012 & 0.6840 & 0.9809 & -1.2881 & -0.9045 & 0.006199\\
M053 & 0.1226 & 0.0226 & 0.7998 & 0.8265 & 1.0161 & -1.2593 & -0.3858 & 0.001096\\
M054 & 0.1345 & 0.0216 & 0.8505 & 0.6251 & 0.8535 & -1.2526 & 0.5703 & 0.007438\\
M055 & 0.1298 & 0.0222 & 0.7504 & 0.8170 & 0.9574 & -1.0573 & 1.0338 & 0.006843\\
M056 & 0.1529 & 0.0219 & 0.8508 & 0.6438 & 1.0322 & -0.7359 & 0.6931 & 0.006311\\
M057 & 0.1419 & 0.0234 & 0.7937 & 0.7415 & 1.0016 & -0.7710 & -1.5964 & 0.005128\\
M058 & 0.1226 & 0.0224 & 0.7278 & 0.6152 & 1.0348 & -1.1051 & 0.2955 & 0.007280\\
M059 & 0.1529 & 0.0224 & 0.7035 & 0.6877 & 0.8616 & -0.9833 & -1.1788 & 0.009885\\
M060 & 0.1270 & 0.0233 & 0.8827 & 0.5622 & 0.8609 & -1.1714 & 0.8346 & 0.009901\\
M061 & 0.1272 & 0.0220 & 0.8021 & 0.8302 & 0.8968 & -0.9545 & -0.6659 & 0.004782\\
M062 & 0.1257 & 0.0216 & 0.7699 & 0.5813 & 0.9460 & -0.8041 & 0.7956 & 0.003922\\
M063 & 0.1312 & 0.0229 & 0.7974 & 0.5890 & 0.9522 & -0.9560 & 0.6650 & 0.001740\\
M064 & 0.1437 & 0.0218 & 0.8866 & 0.7402 & 0.9335 & -1.0713 & 0.7128 & 0.003782\\
M065 & 0.1445 & 0.0218 & 0.8797 & 0.5554 & 0.9353 & -1.2423 & -1.3032 & 0.008850\\
M066 & 0.1392 & 0.0229 & 0.8849 & 0.8176 & 0.8579 & -1.2334 & 0.7098 & 0.000187\\
M067 & 0.1398 & 0.0224 & 0.7795 & 0.7473 & 1.0392 & -1.0471 & 0.7377 & 0.008256\\
M068 & 0.1319 & 0.0232 & 0.7645 & 0.8112 & 0.9199 & -0.7812 & 0.1646 & 0.003716\\
M069 & 0.1392 & 0.0227 & 0.8130 & 0.6062 & 0.8765 & -1.0792 & 0.8817 & 0.006372\\
M070 & 0.1499 & 0.0232 & 0.8093 & 0.7587 & 0.9260 & -0.8942 & -0.9967 & 0.009760\\
M071 & 0.1218 & 0.0224 & 0.7348 & 0.7999 & 1.0330 & -0.9341 & 0.8896 & 0.002212\\
M072 & 0.1265 & 0.0218 & 0.8349 & 0.6080 & 0.9291 & -1.1293 & 0.2197 & 0.002806\\
M073 & 0.1212 & 0.0227 & 0.7683 & 0.6587 & 0.8704 & -0.9662 & 0.9453 & 0.000327\\
M074 & 0.1337 & 0.0216 & 0.8575 & 0.8099 & 0.8518 & -1.0815 & 1.0506 & 0.002370\\
M075 & 0.1484 & 0.0230 & 0.8491 & 0.7212 & 0.9566 & -0.8980 & 0.8181 & 0.002716\\
M076 & 0.1419 & 0.0215 & 0.7700 & 0.6091 & 0.9332 & -0.9753 & -0.2691 & 0.008143\\
M077 & 0.1321 & 0.0230 & 0.7011 & 0.6562 & 0.9938 & -1.1170 & 1.0706 & 0.001979\\
M078 & 0.1446 & 0.0219 & 0.7903 & 0.8074 & 0.9752 & -1.2323 & 1.1681 & 0.004021\\
M079 & 0.1344 & 0.0235 & 0.8742 & 0.7575 & 0.9219 & -1.0483 & -0.3793 & 0.004423\\
M080 & 0.1419 & 0.0218 & 0.8420 & 0.8205 & 0.9145 & -1.1295 & -1.2357 & 0.001959\\
M081 & 0.1366 & 0.0224 & 0.7034 & 0.6599 & 0.8745 & -0.8122 & 0.7675 & 0.005665\\
M082 & 0.1511 & 0.0218 & 0.8061 & 0.7242 & 1.0132 & -0.7940 & 0.0740 & 0.004488\\
M083 & 0.1501 & 0.0230 & 0.7458 & 0.6037 & 0.9999 & -1.2300 & 0.9126 & 0.008023\\
M084 & 0.1244 & 0.0227 & 0.8444 & 0.7462 & 1.0351 & -1.2012 & 1.0042 & 0.002919\\
M085 & 0.1546 & 0.0227 & 0.8201 & 0.8187 & 0.8620 & -1.0794 & 0.3306 & 0.005524\\
M086 & 0.1289 & 0.0221 & 0.8467 & 0.7498 & 0.9158 & -0.8964 & 0.7044 & 0.006605\\
M087 & 0.1437 & 0.0235 & 0.8383 & 0.6612 & 1.0206 & -0.7128 & 0.3537 & 0.006952\\
M088 & 0.1237 & 0.0229 & 0.8779 & 0.6050 & 0.8725 & -1.2333 & 1.1145 & 0.001034\\
M089 & 0.1505 & 0.0221 & 0.8396 & 0.5830 & 0.8506 & -0.8262 & 0.3234 & 0.002603\\
M090 & 0.1439 & 0.0226 & 0.8366 & 0.6987 & 0.9116 & -1.2874 & 0.2509 & 0.009071\\
M091 & 0.1325 & 0.0224 & 0.8076 & 0.6680 & 0.9435 & -0.7361 & -0.9543 & 0.003494\\
M092 & 0.1476 & 0.0215 & 0.8452 & 0.8060 & 0.9855 & -0.9543 & 0.9158 & 0.007598\\
M093 & 0.1389 & 0.0219 & 0.7151 & 0.6105 & 0.9228 & -1.2533 & -0.3018 & 0.004566\\
M094 & 0.1386 & 0.0235 & 0.7699 & 0.7494 & 0.8529 & -1.1243 & 1.0953 & 0.006593\\
M095 & 0.1424 & 0.0223 & 0.8764 & 0.6612 & 0.9422 & -1.2912 & 1.2729 & 0.002028\\
M096 & 0.1404 & 0.0219 & 0.8946 & 0.8155 & 1.0442 & -1.1562 & -0.8081 & 0.001851\\
M097 & 0.1351 & 0.0226 & 0.7560 & 0.6549 & 1.0041 & -0.8389 & 0.8124 & 0.005556\\
M098 & 0.1259 & 0.0225 & 0.7918 & 0.7512 & 0.9055 & -1.1744 & 0.9019 & 0.003027\\
M099 & 0.1237 & 0.0233 & 0.8907 & 0.6375 & 0.9715 & -1.2563 & -0.6293 & 0.007970\\
M100 & 0.1232 & 0.0221 & 0.7715 & 0.5530 & 0.8635 & -0.9174 & -1.7275 & 0.007150\\
M101 & 0.1526 & 0.0217 & 0.7535 & 0.7292 & 0.8836 & -0.7672 & -0.1455 & 0.004596\\
M102 & 0.1531 & 0.0229 & 0.8727 & 0.8137 & 0.9916 & -1.1062 & 0.5227 & 0.005416\\
M103 & 0.1274 & 0.0223 & 0.8993 & 0.7448 & 1.0455 & -0.9231 & 0.7886 & 0.006497\\
M104 & 0.1481 & 0.0222 & 0.7512 & 0.7255 & 1.0029 & -1.0720 & 0.1433 & 0.000910\\
M105 & 0.1490 & 0.0222 & 0.8922 & 0.5780 & 0.9802 & -0.8530 & 0.4716 & 0.002495\\
M106 & 0.1422 & 0.0223 & 0.8679 & 0.6048 & 0.8869 & -0.8012 & 0.6662 & 0.009949\\
M107 & 0.1304 & 0.0233 & 0.8171 & 0.8062 & 1.0495 & -0.8080 & 0.3337 & 0.003607\\
M108 & 0.1414 & 0.0223 & 0.7269 & 0.7524 & 0.9095 & -1.0203 & 0.6065 & 0.008364\\
M109 & 0.1377 & 0.0229 & 0.8656 & 0.6012 & 1.0062 & -1.1060 & 0.9672 & 0.006263\\
M110 & 0.1336 & 0.0220 & 0.8703 & 0.8423 & 0.9509 & -1.1045 & 0.0875 & 0.009305\\
M111 & 0.1212 & 0.0230 & 0.7810 & 0.6912 & 0.9695 & -0.9892 & 0.1224 & 0.007263\\
\tableline
\tableline
\multicolumn{9}{l}{Additional test models}\\
T001 & 0.1333 & 0.02170 & 0.8233 & 0.7444 & 0.9778 & -1.1560 & -1.1220 & 0.005311\\
T002 & 0.1450 & 0.02184 & 0.8078 & 0.6689 & 0.9000 & -0.9333 & -0.5667 & 0.003467\\
T003 & 0.1417 & 0.02300 & 0.7767 & 0.7256 & 0.9222 & -0.8444 & 0.8222 & 0.004389\\
T004 & 0.1317 & 0.02242 & 0.7611 & 0.6878 & 0.9333 & -1.2000 & -0.2889 & 0.001622\\
M000 & 0.1335 & 0.02258 & 0.8 & 0.71 & 0.963 & -1.0 & 0.0 & 0.0
\enddata
\tablecomments{The models are chosen according to a tesselation-based nested design strategy \citep{Heitmann2016ApJ...820..108H}. The models M001--M010 form a complete design with massless neutrinos. Within the main design, the first 26 models (M011--M036), the first 55 models (M011--M066), and the full design (M011--M111) form complete designs, with improving refinement. The emulator presented in this work is constructed using M001--M111.}
\end{deluxetable}

\bibliography{ads}
\bibliographystyle{aasjournal}

\end{document}